\documentclass{ws-ijmpe}
\newcommand{\beq}{\begin{equation}}
\newcommand{\eeq}{\end{equation}}
\newcommand{\beqa}{\begin{eqnarray}}
\newcommand{\eeqa}{\end{eqnarray}}

\newcommand{\Sigs}{\Sigma_{\mathrm s} }
\newcommand{\Sigv}{\Sigma_{\mathrm v} }
\newcommand{\Sigo}{\Sigma_{\mathrm o} }
\newcommand{\kf}{k_{\mathrm F} }

\newcommand{\bfgamma}{\mbox{\boldmath$\gamma$\unboldmath}}
\newcommand{\veck}{\textbf{k}}

\begin{document}

\markboth{H. M\"uther, F. Sammarruca and Zhongyu Ma }{Relativistic Effects and Three-Nucleon Forces                              
}

\catchline{}{}{}{}{}

\title{Relativistic Effects and Three-Nucleon Forces in Nuclear Matter and Nuclei
}

\author{ Herbert M\"uther                                           
}

\address{Institut f\"ur Theoretische Physik, Universit\"at T\"ubingen  \\   D-72076 T\"ubingen, Germany\\  herbert.muether@uni-tuebingen.de                       
} 

\author{Francesca Sammarruca}

\address{Department of Physics, University of Idaho\\ Moscow, Idaho\\ fsammarr@uidaho.edu}

\author{Zhongyu Ma}

\address{China Institute of Atomic Energy\\ P.O. Box 275(41), Beijing 102413, China\\mazy12@ciae.ac.cn}

\maketitle

\begin{history}
\received{(received date)}
\revised{(revised date)}
\end{history}

\begin{abstract}
We review a large body of predictions obtained within the framework of relativistic meson theory together with the Dirac-Brueckner-Hartree-Fock approach to nuclear matter and finite nuclei. The success of this method has been largely related to its ability to take into account important three-body effects. Therefore, the overarching theme of this article is the interpretation of the so-called ``Dirac effects" as an effective three-nucleon force. 
We address the equation of state of isospin symmetric and asymmetric nucleonic matter and related issues,      
ranging from proton and neutron density distributions to momentum distributions and short-range correlations. 
A central part of the discussion is devoted to the optical model potential for nucleon-nucleus scattering.
We also take the opportunity to explore similarities and differences with predictions based on the increasingly popular chiral effective field theory. 
\end{abstract}

\section{Introduction}
\label{intro} 

One of the central challenges of theoretical nuclear physics is to describe the properties of nuclear systems in terms
of a realistic nucleon-nucleon ($NN$) interaction. There exist quite a few phenomenological approaches based on effective interaction models, such as the 
Skyrme forces\cite{skyrme1,skyrme2,skyrme3} or relativistic mean field models,\cite{Wal74,releff2,releff3} which reproduce properties like binding energy and
radii of nuclei all across the nuclide chart with very good accuracy. Although these phenomenological descriptions are very helpful and deserving, 
their parameters have been adjusted to describe specific nuclear properties. As a consequence, 
the predictive power of these models is limited as they fail to describe properties of nuclear systems they were not designed for.
Therefore the microscopic approach is not only a scientific challenge but also an important tool to predict the properties
of nuclear systems which are not directly accessible to experiments. These range from the equation of state (EoS) of astrophysical objects like neutron stars or
the explosion of supernovae to the properties of nuclei far outside the valley of stability, which play an important role in the chains of nuclear
reactions leading to stable isotopes in the universe as well as in nuclear reactors.

The first step for a microscopic nuclear structure calculation is the development of a realistic $NN$ interaction. Throughout this article,  
a realistic $NN$ interaction is a model whose parameters have been adjusted to obtain an accurate fit of the $NN$ scattering data at energies below the pion production threshold and the deuteron properties. 

Over the years different kinds of realistic $NN$ interactions have been developed. A very popular approach is based on the assumption of
a local potential with various kinds of spin-isospin operators consistent with general symmetries. The change of sign in the S-wave phase shifts for $NN$ scattering suggests that a local realistic $NN$ interaction should contain repulsive components of short range
and attractive components with longer ranges. In fact, some of the early
models for a local $NN$ interaction described the repulsive components in terms of a hard core potential, with infinite
repulsion for relative distances below $r_c \approx 0.4$ fm\cite{hamada,reid} and strong attraction at larger distances. More modern versions of       
a local $NN$ interaction, such as the Argonne potentials,\cite{wiring84} also contain a considerable amount of cancellation between very
repulsive components of short range and attractive ones at larger ranges. The long-range part of these interactions is dominated by a local approximation
of a one-pion-exchange term.  Although the one-pion-exchange component generates very strong spin- and tensor-components, its contribution to the energy 
of spin saturated nuclear systems is very small.

After the discovery of heavy mesons, the meson exchange or One-Boson-Exchange-Potentials (OBEP) became very 
popular.\cite{erkelenz74,holinde-mach75,nijm78,machl89,nijm94,machl01,brown-jack} Even today the OBEP are the most efficient models of realistic $NN$ interactions in the sense that they yield a fit of the $NN$ scattering data with high accuracy with a minimal number of
parameters to be adjusted. The medium range attraction in the OBEP is mainly due to the exchange of a fictitious scalar isoscalar ``meson", the $\sigma$ or $\epsilon$ meson, which is introduced to parametrize the correlated exchange of two pions. Various attempts have been made to describe the features of this phenomenological $\sigma$ meson in terms of dispersion relations\cite{brown-jack,stony15,paris17,paris18} or by explicit evaluation of non-iterative two-pion-exchange terms or two meson exchange terms with intermediate excitation of the interacting nucleon, in particular to the $\Delta(1232)$ resonance.\cite{machl87,HMAFM78,HMAFM79} The repulsive short-range components of the OBEP are generated through the exchange of vector mesons, in particular the isoscalar $\omega$ meson.

With the success of Quantum Chromo Dynamics (QCD) in describing the properties of hadrons, QCD inspired quark models of the $NN$ interaction were constructed.\cite{harvey81,okayaz81,faeslueb,thom83,takeuch89,entemfern} These models were able to describe the qualitative features of the $NN$ scattering data, but have not been developed to a level of high-accuracy as e.g. within the OBE approach.

During the past decade $NN$ interaction models based on chiral effective field theories (EFT) have become very popular.\cite{kaiser97,kaiser98,bocum00} The chiral theory leads to a perturbative expansion and provides a consistent description of two- and many-nucleon forces. The Idaho group managed to construct a chiral $NN$ potential of high precision at next-to-next-to-next-to-leading order (N$^3$LO), which has been used in few-nucleon and nuclear structure calculations.\cite{n3lo,ME11} 

The construction of a realistic $NN$ interaction, however, is only the first step. In a second step this $NN$ interaction is used in many-body calculations of nuclear systems without any additional parameters to adjust. As a first testing ground one typically tries to evaluate the energy per nucleon in infinite nuclear matter as a function of density. For isospin symmetric nuclear matter one would like to reproduce the so-called saturation point, i.e. the empirical values for the minimum of the energy {\it vs.} density curve, which, according to the Weiz\"acker mass formula, should occur at -16 MeV per nucleon at a nuclear density around 0.16 fm$^{-3}$. Extension of the study to isospin asymmetric nuclear matter allows for the evaluation of the symmetry energy.

All the realistic models of the $NN$ interaction discussed above show very strong attractive and repulsive components, so that a simple mean-field description or a perturbative treatment based on the the mean field model will not work. Note that the corresponding matrix elements of local hard core potentials will even diverge. The Hartree-Fock approach leads to unbound nuclear systems also for other realistic $NN$ interaction models.\cite{polls2000} The cancellation of attraction and repulsion in the central part and also the strong tensor components originating from the $\pi$-exchange induce two-nucleon correlations which must be treated in a non-perturbative way. One possibility of doing so is the so-called Brueckner-Hartree-Fock (BHF) approximation. The key equation of this approach is the Bethe-Goldstone equation,
\begin{equation}
G(\omega )  = V + V \frac{Q}{\omega - H_0} G\,,\label{eq:bg}
\end{equation}
which defines an effective interaction $G$ in terms of the bare $NN$ interaction $V$. The $G$ matrix is formally very similar to the Lippman-Schwinger equation for the $T$ matrix, which describes the $NN$ interaction in vacuum. The main differences are: 1) the intermediate two-particle states are restricted by the Pauli operator, $Q$, to unoccupied two-nucleon states, namely with momenta larger than the Fermi momentum; and 2) $\omega$ and $H_0$ in the energy denominator are defined in terms of the single-particle energies $\varepsilon_i$ of the nuclear many-body system. In this sense the Bethe-Goldstone equation, Eq.~(\ref{eq:bg}), represents the solution of the two-nucleon problem imbedded in the nuclear medium and accounts for two-nucleon correlations and the medium dependence of these correlations. The simplest approach to define these energies is the Hartree-Fock prescription of replacing the bare interaction $V$ by the matrix elements of $G$, 
\begin{equation}
\varepsilon_i = t_i + \sum_{j<F} \langle ij\vert G(\omega = \varepsilon_i + \varepsilon_j) \vert ij\rangle\rho_j \,,
\label{eq:bhf1}
\end{equation}
where $t_i$ denotes the kinetic energy for a nucleon in state $i$ and the sum is restricted to states $j$ below the Fermi energy, which is taken into account by the single-particle density $\rho_j$.
It is clear that an iterative procedure is required to obain a self-consistent solution of Eqs.~(\ref{eq:bg}) and (\ref{eq:bhf1}). The total energy can then be calculated using
\begin{equation}
E = \sum_{i<F} t_i + \frac12\sum_{i,j<F} \langle ij\vert G(\omega = \varepsilon_i + \varepsilon_j) \vert ij\rangle\rho_i\rho_j = \frac12\sum_{i<F}(t_i + \varepsilon_i)\,.
\label{eq:koltun}
\end{equation}
In contrast to the Hartree-Fock approach such BHF calculations for nuclear matter yield a saturation point, i.e. a minimum in the energy per nucleon {\it vs.} density curve with negative energy. At this point one might argue that it is sufficient to ignore the medium dependence of the correlation effects and replace the $G$ matrix by the corresponding $T$ matrix. It turns out, however, that such $T$ matrix approach yields too much binding energy and a saturation density which is very large. So it is important to account for the quenching of the attractive higher order terms in $G$, which is due to the Pauli operator (Pauli quenching) and the larger energy denominator (dispersive quenching). 

It turns out that BHF calculations using realistic $NN$ potentials yield saturation points for nuclear matter which are all positioned in a small band in the energy {\it vs.} density plane, the so-called Coester band,\cite{coester,polls2000} which does not meet the empirical point. In fact, the position of a specific $NN$ potential along this Coester band depends on the strength of the Pauli- and dispersive quenching effects. Stiff potential with strong short-range components and/or a strong tensor force generate large terms of higher order in $V$ in the $G$-matrix. In turn, these lead to strong quenching effects and, therefore, small binding energies. Softer $NN$ potentials, on the other hand, show weaker quenching effects and predict to much binding energy as compared to the empirical value at a saturation density which is twice as large or even larger than the empirical density. The $D$-state probability in the deuteron is a measure for the strength of tensor correlation effects and therefore one indicator for the position in the Coester band: A small $D$-state probability corresponds to a soft interaction with a saturation point at large binding energy and density.

The BHF approximation is of course only a very simple approximation for the solution of the nuclear many-body problem. Many attempts have been made to improve the many-body theory either within the hole-line expansion by including the effects of three-nucleon ($3N$) correlations\cite{day81,song98} or considering different approximation schemes. These include variational calculations using correlated basis functions,\cite{fabrocini} Quantum Monte Carlo calculations,\cite{pieper} the coupled cluster or exponentional S method,\cite{kuemmel,mihaila} and the self-consistent Green's function method.\cite{frick,arturox} These improvements and variations in the many-body approach lead to modifications in the details of the results but do not change the basic features of the BHF approximation: Nonrelativistic calculations using realistic two-nucleon interactions cannot reproduce the empirical saturation point of nuclear matter. The calculated saturation point lays on the Coester band depending on the softness of the interaction.

Therefore, one has to introduce three-nucleon forces (3NF) (or, more generally, many-nucleon forces) to obtain the empirical saturation point of nuclear matter within a non-relativistic theory based on realistic $NN$ interactions. Important 3NF are implied by the underlying theory of the strong interaction. For example, 3NF occur in the chiral EFT mentioned earlier or from the consideration of subnucleonic degrees of freedom through the inclusion of baryon excitations. These lead to many-nucleon forces if the corresponding degrees of freedom are eliminated from the Hilbert space explicitly considered. A successful description of the empirical saturation point, however, could only be obtained by considering phenomenological 3NF with adjustable parameters.\cite{urbana3,baldofit,ericfit}
 
On the other hand, relativistic calculations have been successful in describing the empirical saturation point of symmetric nuclear matter from realistic OBEP without introducing phenomenological 3NF.\cite{An83,HS84,BM84,HM87,AS03,Serot1,poschenr,jongl98,schiller,dalen1,dalen2,dalenrev}  
At first sight relativistic effects seem to be negligible
when dealing with the nuclear many-body problem, as the binding energies of nucleons
are very small as compared to the nucleon rest mass. The meson exchange
model of the $NN$ interaction, however, predicts a strong cancellation of attractive and repulsive components
in the relativistic self-energy of nucleons in a nuclear medium. A strong
attractive part (approximately -300 MeV), which transforms like a scalar under a Lorentz
transformation and mainly originates from the exchange of the scalar $\sigma$ meson in
realistic OBE models, is to a large extent compensated
by a repulsive component, which transforms like a time-like component
of a Lorentz vector and reflects the repulsion due to the $\omega$ meson exchange. The resulting single-particle energy of the nucleon ends up to be approximately -40 MeV, which is indeed small on the
scale of the nucleon mass.                  

Inserting the nucleon self-energy in a Dirac equation for the nucleon, the strong
scalar part of the self-energy leads to  an enhancement of the small component in
the nucleon Dirac spinor in the medium as compared to the vacuum. This     
modification of the Dirac spinor yields modified matrix
elements for the OBE interaction. It is this density dependence of the nucleon Dirac
spinor and the resulting medium dependence of the $NN$ interaction which moves
the saturation point in nuclear matter off the Coester band  and to its empirical value. 
 
Studies of infinite matter are quite insightful, but only probe the 
volume effects for the bulk properties of nuclear systems. Various attempts have been made to apply the DBHF approach also to the description of finite nuclei.
Due to the additional complications encountered in the description of finite nuclei, most of these attempts are based on DBHF studies of nuclear matter together with a Local Density Approximation (LDA) to describe the ground state properties of finite nuclei.\cite{muth:1987,fritz:1993,fritz:1994,ulrych:1999} Only very recently a DBHF calculation has been reported which avoids these approximations, solves the DBHF equations directly for $^{16}$O and indeed obtains a very good description of the total energy and radius of this nucleus without any adjustable parameters.\cite{ring2016} 

Therefore the time seems appropriate to review the DBHF approach\cite{dalenrev} beyond its applications to infinite matter and focus also on DBHF predictions of other ``observables". After a brief review of the DBHF scheme in Section~\ref{sect2}, we will discuss some issues of contemporary interest with regard to isospin-asymmetric infinite matter and the closely related symmetry energy. These include the relation between neutron skins in neutron-rich nuclei and the pressure in neutron matter (Section~\ref{sect31}), polarization in asymmetric matter (Section \ref{sect32}), and momentum distributions due to correlation effects (Section~\ref{sect33}). 
 When appropriate, we will take the opportunity to compare DBHF predictions with those obtained with other modern descriptions of nuclear systems, the chiral EFT in particular.
 The Optical Model Potential for nucleon - nucleus scattering based on DBHF will be discussed in Section \ref{sect4} and attempts to describe  relativistic features of DBHF in terms of 3NF will be reported in Section \ref{sect5}.  We will present a brief summary and conclusions in Section \ref{sect6}.                                                                                                                                                                                                            

\section{General aspects of the DBHF approach\label{sect2}}
\label{dbhf_general}

The self-energy $\Sigma$ is the single-particle operator describing the interaction of a nucleon with the surrounding medium. For a nucleon in infinite nuclear matter, its structure follows from translational and rotational invariance,
hermiticity, parity conservation, and time reversal invariance.                    
Therefore, the most general form of the Lorentz structure of the self-energy  is given by
\beqa
\Sigma = \Sigma_s - \gamma_\mu \Sigma^\mu
\label{subsec:DBHF;eq:selfour1}
\eeqa
with
\beq
\Sigma^\mu = \Sigo u^\mu + \Sigv  \Delta^{\mu\nu}k_\nu,
\label{subsec:DBHF;eq:selfour2}
\eeq
where the $\Sigs$, $\Sigo$, and $\Sigv$ components are Lorentz scalar
functions, which depend on the Lorentz invariants $k^2$, $k
\cdot  j$ and $j^2$, with $j_{\mu}$ the baryon current and $k_{\mu}$ the nucleon four-momentum.
Therefore, these Lorentz invariants can be expressed in terms of
$k_0$, $|\veck|$, and the Fermi momentum $\kf$.
Furthermore, the projector $\Delta^{\mu\nu}$ in Eq.~(\ref{subsec:DBHF;eq:selfour2}) is given by
$\Delta^{\mu\nu} = g^{\mu\nu} - u^\mu u^\nu$ with streaming velocity $u^\mu = j^\mu / \sqrt{j^2}$.
In the nuclear matter rest frame, where $u^{\mu} = \delta^{\mu 0}$, the self-energy then has the simple form
\beq
\Sigma(k,\kf)= \Sigs (k,\kf) -\gamma_0 \, \Sigo (k,\kf) +
\bfgamma  \cdot \textbf{k} \,\Sigv (k,\kf).
\label{subsec:SM;eq:self1}
\eeq
Inserting  this self-energy into the Dirac equation for the nucleon one can rewrite the equation in the form 
\begin{equation}
\left[ k^* \!\!\!\!\!\! / - m^* -i \, \Im m\, \Sigma \right] u(k) =0.
\label{dirac}
\end{equation}
The imaginary contribution to the self-energy, $\Im m\, \Sigma$, is due to the possible decay of particle states with energies above the Fermi energy. It will become relevant when we discuss the structure of the optical potential for nucleon scattering in Section \ref{sect4} but will be ignored in the present context. We note that the impact of the self-energy in the Dirac equation can be interpreted as a modification of the bare mass $M$ of the nucleon to an effective mass as well as a modification of the four momentum $k^\mu$ in the nuclear medium. More precisely, 
\beqa
m^*(k, \kf)  = M + \Re e \Sigma_s(k, \kf), \quad k^{*\mu}=k^{\mu} + \Re e
\Sigma^{\mu}(k, \kf).
\label{subsec:SM;eq:dirac}
\eeqa
Introducing the reduced effective mass,
\beqa
{\tilde m}^*(k,\kf) = m^*(k,\kf)/ \left( 1+\Sigv(k,\kf)\right),
\label{subsec:SM;eq:redquantity}
\eeqa
and the reduced kinetic momentum,
\beqa
{{\tilde k}^*}_\mu = k^*_\mu / \left( 1+\Sigv(k,\kf)\right),
\eeqa
the Dirac equation written in terms of these reduced effective quantities 
takes the form
\beqa
 [ \gamma_\mu {\tilde k}^{*^\mu} - {\tilde m}^*(k,\kf)] u(k,\kf)=0.
\quad
\label{subsec:SM;eq:dirac2}
\eeqa
The solution of the Dirac equation written as in Eq.~(\ref{subsec:SM;eq:dirac2})  is
\beqa
u_\lambda (k,\kf)= \sqrt{ \frac{ {\tilde E}^*(\veck)+ {\tilde m}^*_F}
{2{\tilde m}^*_F}}
\left(
\begin{array}{c} 1 \\
\frac{2\lambda |\veck|}{{\tilde E}^*(\veck)+ {\tilde m}^*_F}
\end{array}
\right)
\chi_\lambda,
\label{spinor}
\eeqa
where ${\tilde E}^*(\veck)=\sqrt{\veck^2+{\tilde m}^{*2}_F}$ denotes the reduced effective energy
and $\chi_\lambda$ a two-component Pauli spinor with
$\lambda=\pm {\frac{1}{2}}$.
The normalization of the Dirac spinor is
thereby chosen as $\bar{u}_\lambda(k,\kf) u_\lambda(k,\kf)=1$.
From Eq.~(\ref{subsec:SM;eq:dirac2}) one derives the single-particle potential ${\hat U } = \gamma^0  \Sigma$. It can be obtained by calculating the expectation value of ${\hat U}$. Therefore, one sandwiches ${\hat U}$ between
the effective spinor basis, Eq.~(\ref{spinor}),
\begin{eqnarray}
   U(k) = \frac{\langle u(k)|\gamma^0  \Sigma | u(k)\rangle }
{\langle u(k)| u(k)\rangle } =
\frac{{\tilde m}^*}{{\tilde E}^*(\veck)}
\, \langle {\bar u(k)}| \Sigma | u(k)\rangle = \frac{{\tilde m}^*}{{\tilde E}^*(\veck)} {\tilde \Sigs} - {\tilde \Sigo}.
\label{upot1}
\end{eqnarray}
Using a realistic OBE model for the $NN$ interaction one obtains a very attractive scalar component $ {\tilde \Sigs}$ of the order of  -300 MeV around saturation density of
nuclear matter, which has its main origin in the exchange of the $\sigma$ meson, the scalar-isoscalar meson responsible for the medium range attraction. When calculating the single-particle potential $U$, this attraction is compensated to a large extent by the repulsion due to the time-like vector component in the self-energy, $\tilde \Sigo$ and the resulting value of the single-particle potential is small as compared to the bare mass of the nucleon $M$, which would suggest small relativistic effects.

The large scalar component, however, yields a low effective mass ${\tilde m}^*$ (see Eq.~(\ref{subsec:SM;eq:redquantity}) and (\ref{subsec:SM;eq:dirac})) and consequently a substantial enhancement of the lower component in the Dirac spinor of the nucleon in the medium (see Eq.~(\ref{spinor})). Using these Dirac spinors with enhanced as well as density dependent small component generates a repulsive effect which shifts the predicted saturation point of symmetric nuclear matter towards the empirical value.

This characteristic feature originating from the medium dependence of the nucleon Dirac spinors, which is essential for the saturation of nuclear matter, is already contained in the simple mean-field model of the Walecka type.\cite{Wal74,Serot1} Within the DBHF approach, however, one adopts a realistic OBEP and solves a Bethe-Goldstone type of equation (see Eq.~(\ref{eq:bg}), the Thompson equation\cite{Thom} in particular, which is a relativistic three-dimensional reduction of
the Bethe-Salpeter equation.\cite{BS}
The various components of the self-energy are then determined either by a fitting procedure for the single-particle spectrum\cite{BM84} or a projection technique, which allows for a more detailed analysis of the Dirac components of $\Sigma$.\cite{HM87,dalen1} In any case, the DBHF approach yields the empirical saturation point of nuclear matter without adjustment of any parameter due to the Dirac effects described above and the quenching of the $NN$ correlation discussed in the Introduction.

\begin{figure}[!t] 
\centering         
\vspace*{-3.2cm}
\hspace*{-0.5cm}
\scalebox{0.9}{\includegraphics{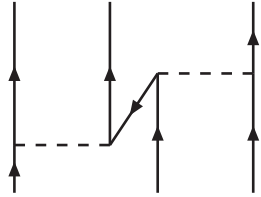}}
\vspace*{-19.0cm}
\caption{Three-body force due to virtual pair excitation.                                          
} 
\label{3b}
\end{figure}

The enhancement of the small component of the nucleon Dirac spinor in the medium can be understood as a virtual excitation of a nucleon-antinucleon pair due to the interaction with the other nucleons. This is represented by the well known ``Z-diagram'' shown in Fig.~\ref{3b}, which helps visualize how relativistic effects can be interpreted in terms of a 3NF. It was estimated by G.E. Brown\cite{GB87} that the enhancement of the small component or lowering of the effective Dirac mass  $\tilde m^*$ as well as the evaluation of the Z-diagram, Fig.~\ref{3b}, generate a repulsive effect on the energy per particle in symmetric
nuclear matter which depends on the density approximately as 
\begin{equation}
\Delta E \propto  \left (\frac{\rho}{\rho_0}\right )^{8/3} \, , 
\label{delE} 
\end {equation}
and provides the saturating mechanism missing in conventional Brueckner-Hartree-Fock (BHF) calculations. Alternatively, explicit 3NF can be used along with the BHF method in order to achieve a similar result.

Various attempts have been made to parameterize the 3NF discussed so far in terms of
a simple local 3NF. As an example we mention the Urbana force,\cite{urbana3,baldofit} which is composed
of two terms
\begin{equation}
V_{ijk} = A \,V_{ijk}^{2\pi} + U\,V_{ijk}^R\,.\label{eq:3-v}
\end{equation}
The first part is from $2\pi$ exchange with an intermediate $\Delta$ excitation and may 
be regarded as simulating the medium-dependence of subnuclear degrees of freedom. The second
term is typically defined in terms of $2\sigma$ exchange and can be interpreted  as a way to simulate
the effects of the $Z$-diagram discussed above. In other words, the second term can be thought to       
represent the relativistic effects of the DBHF approach.\cite{grange} Typically this 3NF 
is reduced to a density-dependent $NN$ interaction, which is then added to the bare $NN$ interaction
(see e.g. \cite{soma} and references therein). The parameters $A$ and $U$ in Eq.~(\ref{eq:3-v})
can be adjusted to reproduce the empirical saturation point for symmetric nuclear matter.

This scheme has been criticized by Hebeler and Schwenk\cite{hebschw} and later by Carbone, Rios, and Polls.\cite{carbone}
They argue that an expression for the total energy with kinetic energy $t_i$, $NN$ interaction $V_{ij}$, and
$3N$ potential $V_{ijk}$, 
\begin{equation}
E = \sum_i t_i\rho_i + \frac{1}{2}\sum_{i,j} V_{ij}\rho_i \rho_j + \frac{1}{6} \sum_{i,j,k} V_{ikj}\rho_i \rho_k \rho_j \;, \label{eq:3n1}
\end{equation}
leads to the single-particle energy
\begin{equation}
\varepsilon_i = t_i + \sum_j V_{ij}\rho_j + \frac{1}{2} \sum_{j,k} V_{ikj}\rho_k \rho_j\,,\label{eq:3n2}
\end{equation}
which is different from the result obtained when the 3NF is added to the $NN$ interaction as 
\begin{equation}
\frac{1}{2} V_{ij}^{eff}(\rho )  = \frac{1}{2} V_{ij} + \frac{1}{6}\sum_k V_{ikj}\rho_k \; .\label{eq:v3dens}
\end{equation}
This is of course true and at first sight it would imply that the medium effects discussed above would
lead to different results when they are treated as a 3NF or as a density-dependent
$NN$ interaction. We note, however, that both approaches lead to the same result if the single-particle
energies are defined according to the Landau definition of the quasiparticle energy, i.e.
\begin{equation}
\varepsilon_i = \frac{\partial}{\partial \rho_i} E(\rho)\,,\label{eq:landaud}
\end{equation}
which means that rearrangement terms due to the density dependence of $V^{eff}$ are taken into account. With this
inclusion, the result is the same whether the medium effects are treated as a 3NF or a density-dependent $NN$ contribution. Note that consideration
of 3NF, as well as the inclusion of rearrangement terms in the single-particle potential, invalidate the Koltun sum rule for the
energy, which is the second equality in Eq.~(\ref{eq:koltun}).
We will return to the treatment of relativistic effects in terms of a 3NF\cite{lippok} in Section \ref{sect5} below.

\begin{figure}[htb]
\begin{center}
\hspace*{-2.0cm}
\includegraphics[width=7.0cm]{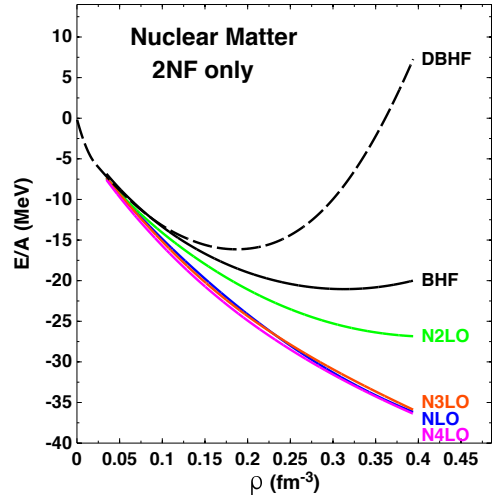}
\vspace*{-0.1cm}
\caption{(Color online) Energy per nucleon in symmetric nuclear matter as a function of density $\rho$.
The DBHF prediction is the
dashed black curve, while the solid black is obtained in a conventional BHF calculation. The blue, green, red, 
and purple curves show EFT-based predictions obtained with only 2NF at 
NLO, N$^2$LO, N$^3$LO, and N$^4$LO, respectively.  } 
\label{ea}
\end{center}
\end{figure}

\section{Predictions for equations of state and related properties:
DBHF {\it vs.} chiral EFT\label{sect3}} 
\label{eos} 

In this section, we will show basic predictions for the EoS in symmetric nuclear matter (SNM) and
neutron matter (NM) obtained through procedures consistent with those described previously, and discuss them 
in comparison with alternative approaches.

As discussed in Section~\ref{intro}, the successful description of nuclear matter saturation through DBHF
is based on the quenching of the attractive two-nucleon correlation effects in the Brueckner $G$-matrix, 
as well as the relativistic feature of reduction of the Dirac mass in the nuclear medium.
This effect can be seen from the energy versus density plot for symmetric nuclear matter in Fig.~\ref{ea}. The two-body interaction part (2NF) obtained in chiral Effective Field Theory  (NLO, N$^3$LO and N$^4$LO) can be considered as very soft descriptions of the $NN$ interaction. Therefore $NN$ correlations obtained within these approaches are very
small and the quenching of these correlation effects in the nuclear medium is almost negligible. Therefore the blue, red and purple curves in Fig.~\ref{ea} almost coincide
and do not show any saturation feature. The Bonn B potential\cite{machl89} is a bit ``stiffer", in particular it contains a stronger tensor component due to the $\pi$-exchange part. Therefore a BHF calculation (black solid line) yields a minimum in the energy versus density plot.\cite{Sam11} As for other realistic $NN$ interactions, the density and energy at the minimum are not realistic . Only with inclusion of the relativistic effects of the DBHF approach (dashed line in Fig.~\ref{ea}) the saturation point shifts towards the empirical value.

\begin{figure}[!htb]
\includegraphics[width=6.0cm]{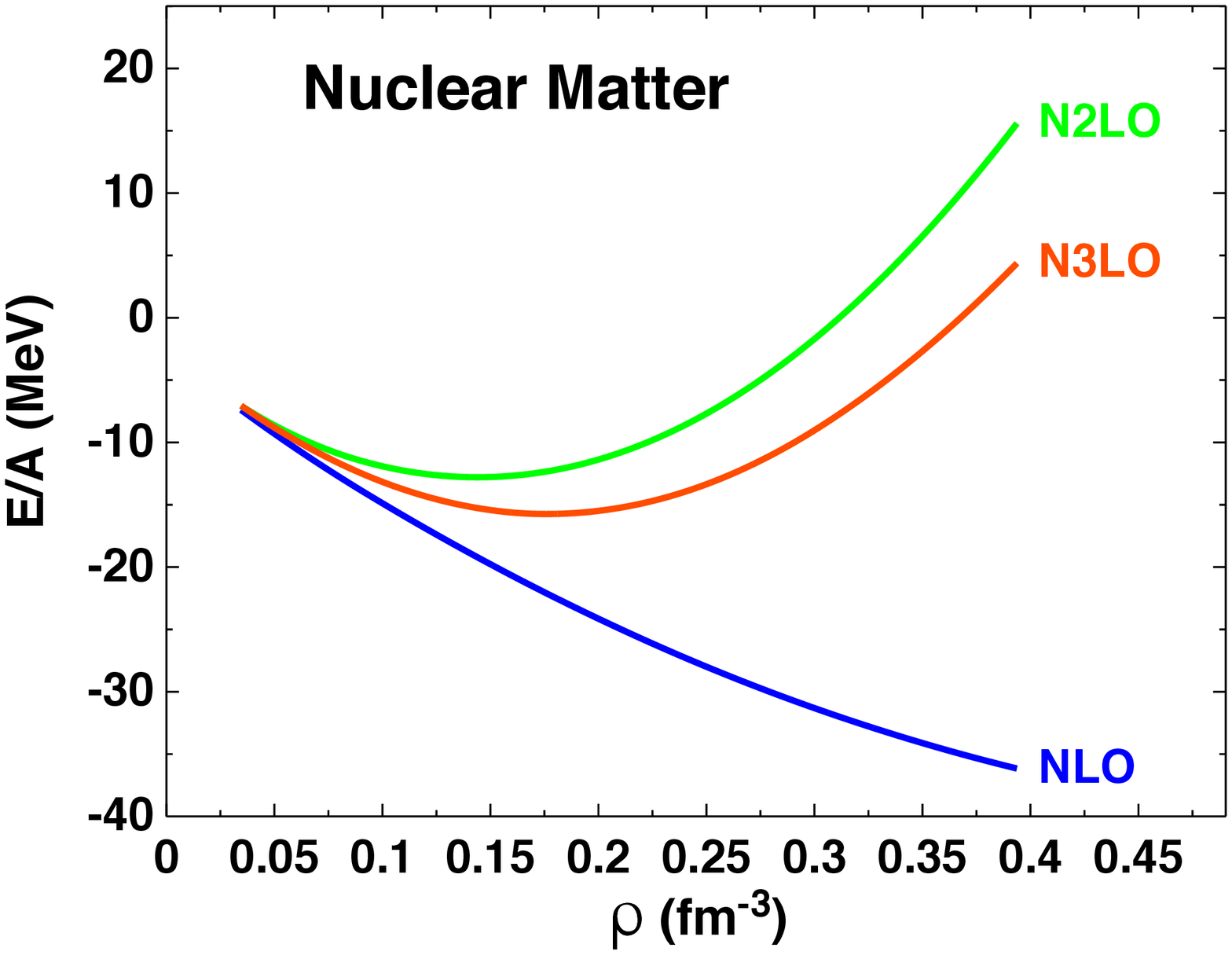}
\includegraphics[width=6.0cm]{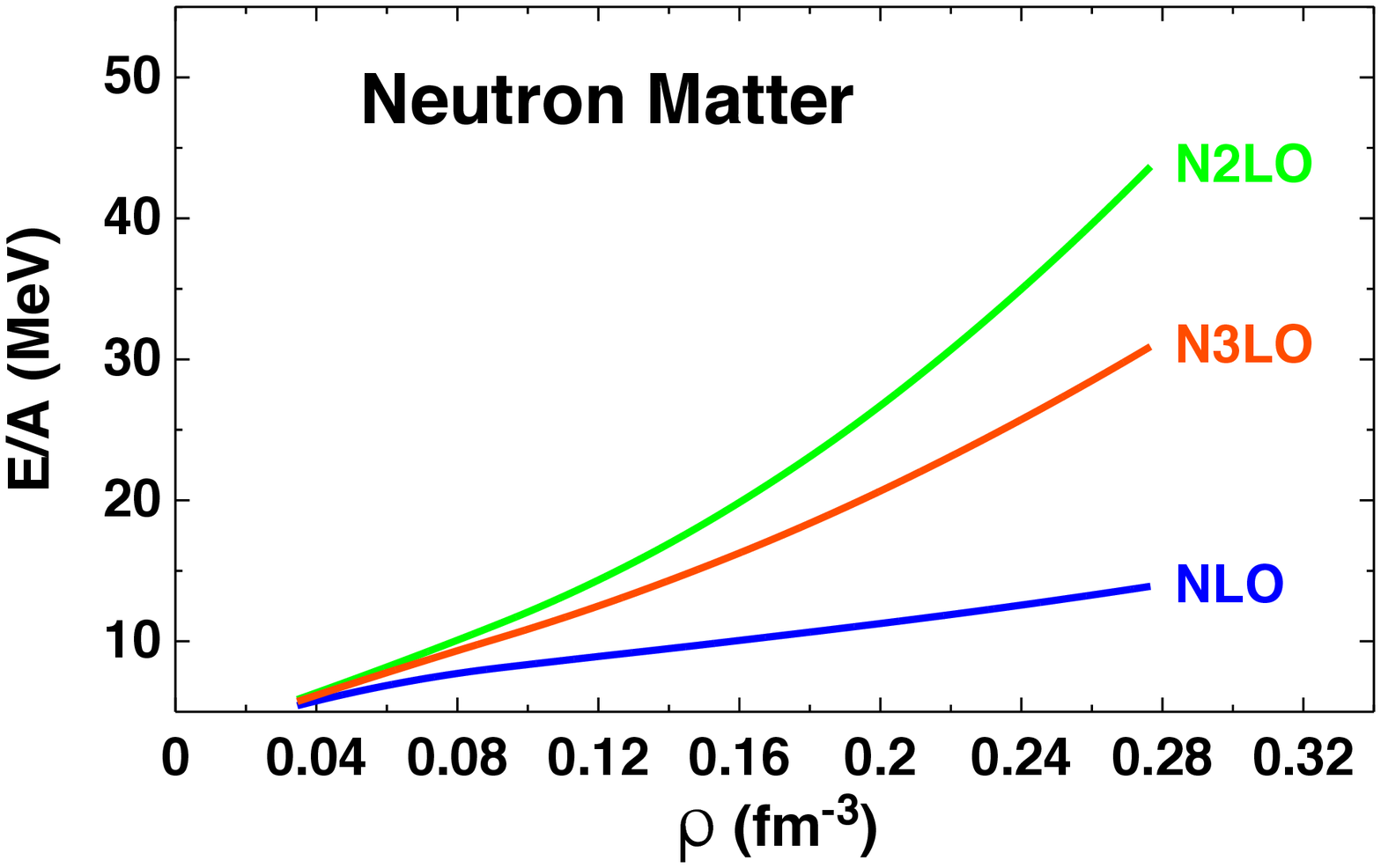}
\vspace*{-0.5cm}
\caption{(Color online) Left frame: Energy per nucleon in symmetric nuclear matter
as a function of density with increasing order of chiral EFT. Right frame: Same as left frame but for 
neutron matter.} 
\label{snm_nm}
\end{figure}

In line with the goals announced earlier in the article, we also show predictions based on
chiral EFT, see Fig.~\ref{snm_nm}. The three curves are obtained at next-to-leading order 
(NLO, or second order), next-to-next-to-leading order (N$^2$LO or third order) and 
next-to-next-to-next-to-leading order (N$^3$LO or fourth order) of the chiral expansion, as described
in Ref.\cite{obo}. The NN potentials are taken from Refs.\cite{nlo,n3lo,ME11}, using 
a cutoff of 450 MeV in the regulator function. A detailed description of the chiral two- and few-nucleon 
forces which we employ can be found in Ref.\cite{obo}

Although a discussion of chiral EFT is outside the purview of this article, we briefly mention some of its     
main features, to render our comparison more insightful.                                  
In EFT, long-range physics is determined by the interaction of pions 
and nucleons constrained by the (broken) symmetries of QCD, whereas 
short-range physics is included through the processes of regularization and renormalization.
Furthermore, at each order of chiral perturbation theory ($\chi$PT), the uncertainty associated 
with a particular prediction can be controlled and quantified.                                     
Applying an organizational scheme to rank-order the various contributions, known as power counting, 
two- and few-nucleon forces emerge on an equal footing in a well-defined hierarchy.                               

\begin{figure}
\begin{center}
\vspace*{-0.5cm}
\includegraphics[width=7.0cm]{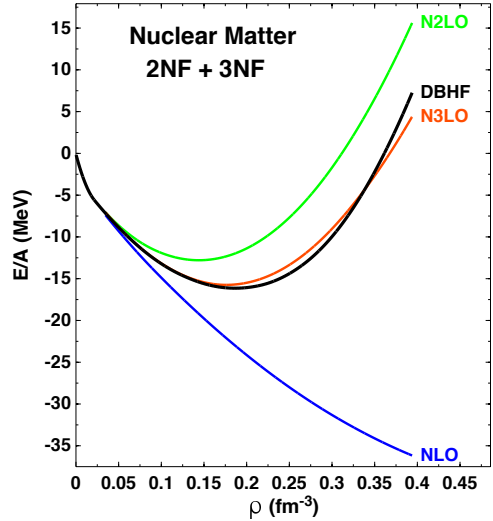}
\vspace*{-0.1cm}
\caption{Energy per nucleon in symmetric nuclear matter as a function of density.
 The solid black curve is the DBHF prediction, whereas the blue, green, and red curves display
the EFT-based predictions at NLO, N$^2$LO, and N$^3$LO, respectively.                                         
} 
\label{snm_dbhf}
\end{center}
\end{figure}

\begin{figure}
\begin{center}
\vspace*{-1.0cm}
\includegraphics[width=6.0cm]{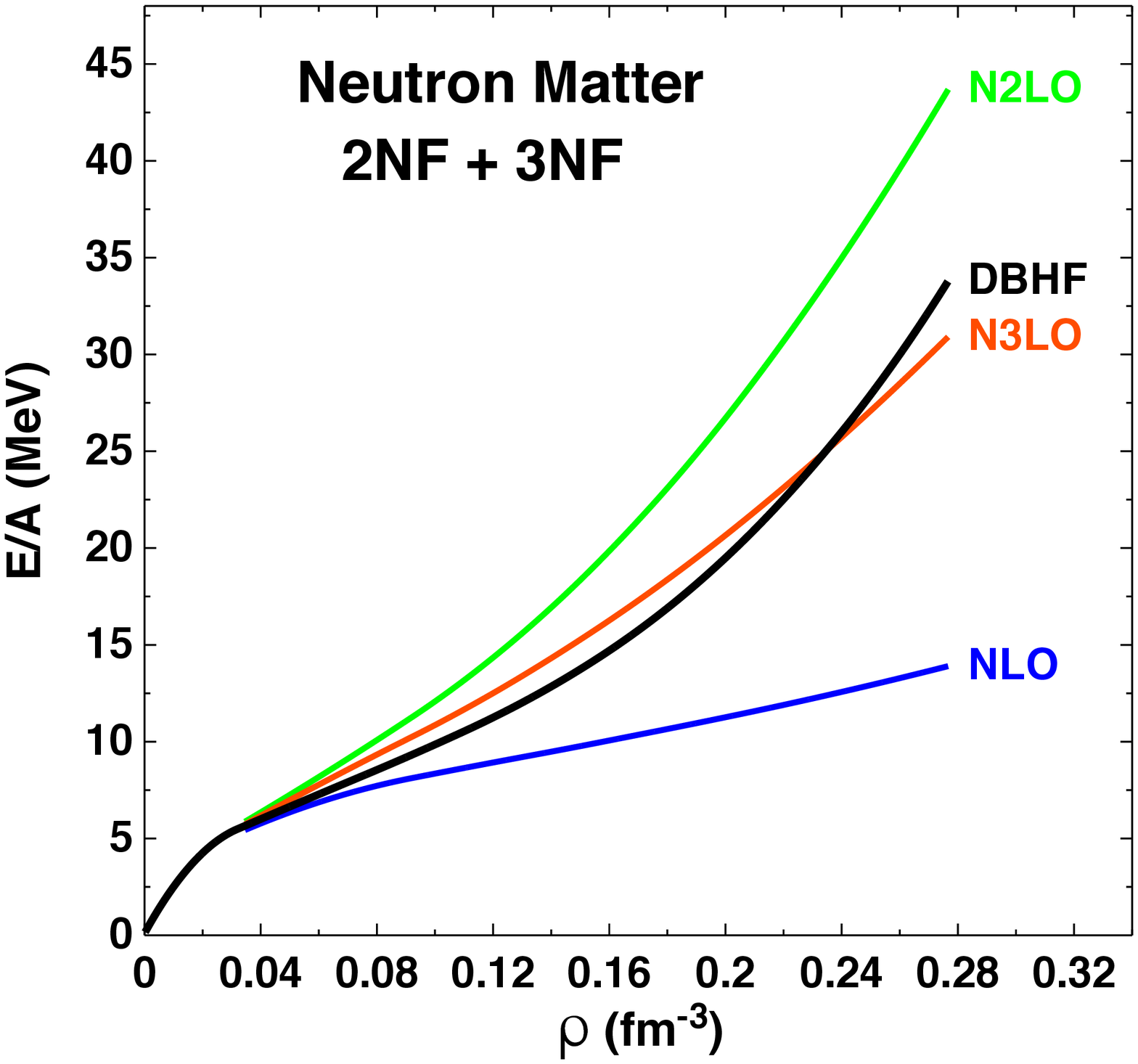}
\includegraphics[width=6.0cm]{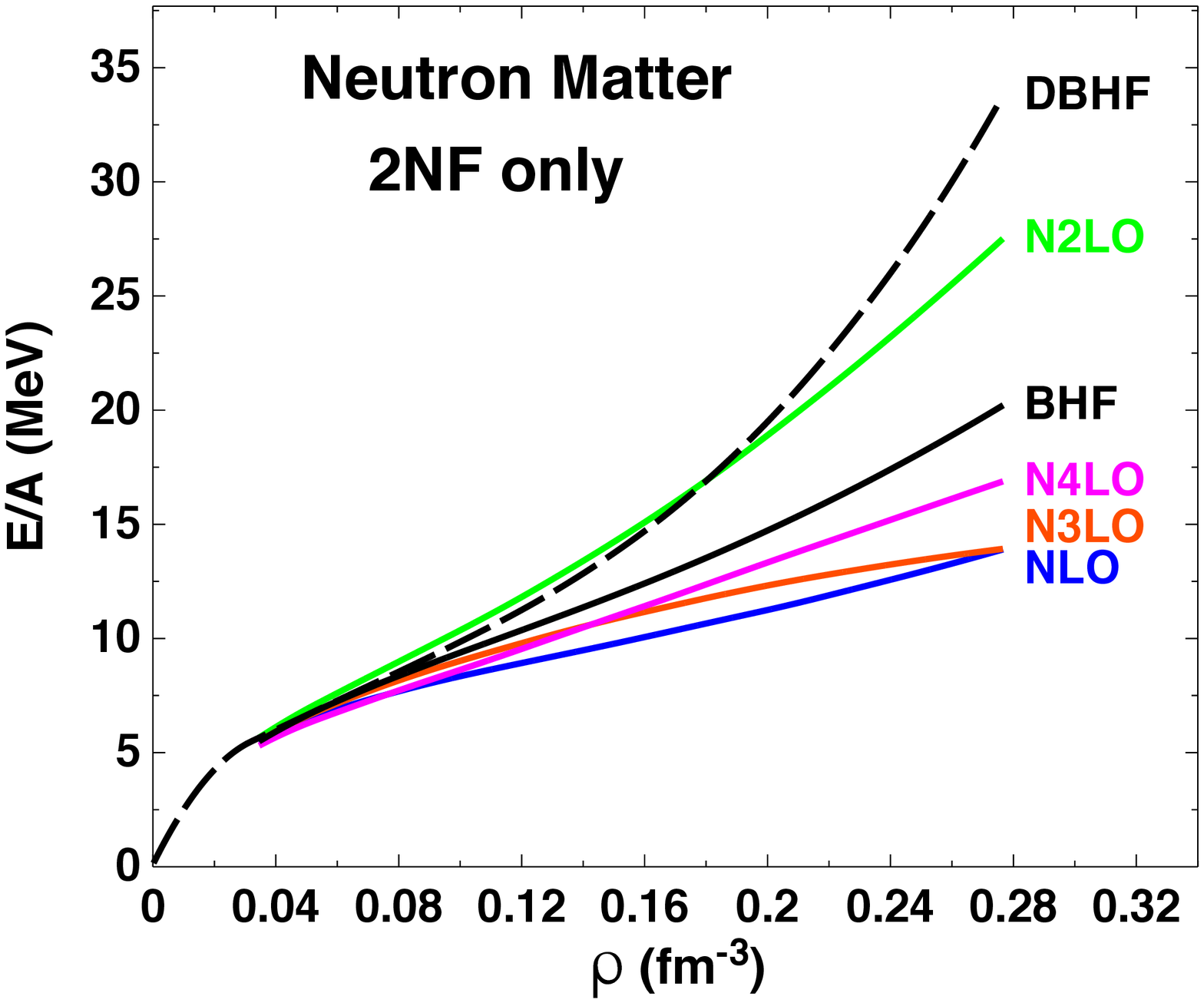}
\vspace*{-0.1cm}
\caption{Same as Fig.~\ref{snm_dbhf} (left) and Fig.~\ref{ea} (right) but for neutron matter.                           
} 
\label{nm_dbhf}
\end{center}
\end{figure}

\begin{figure}[!t]
\hspace*{1.5cm}
\includegraphics[width=8.0cm]{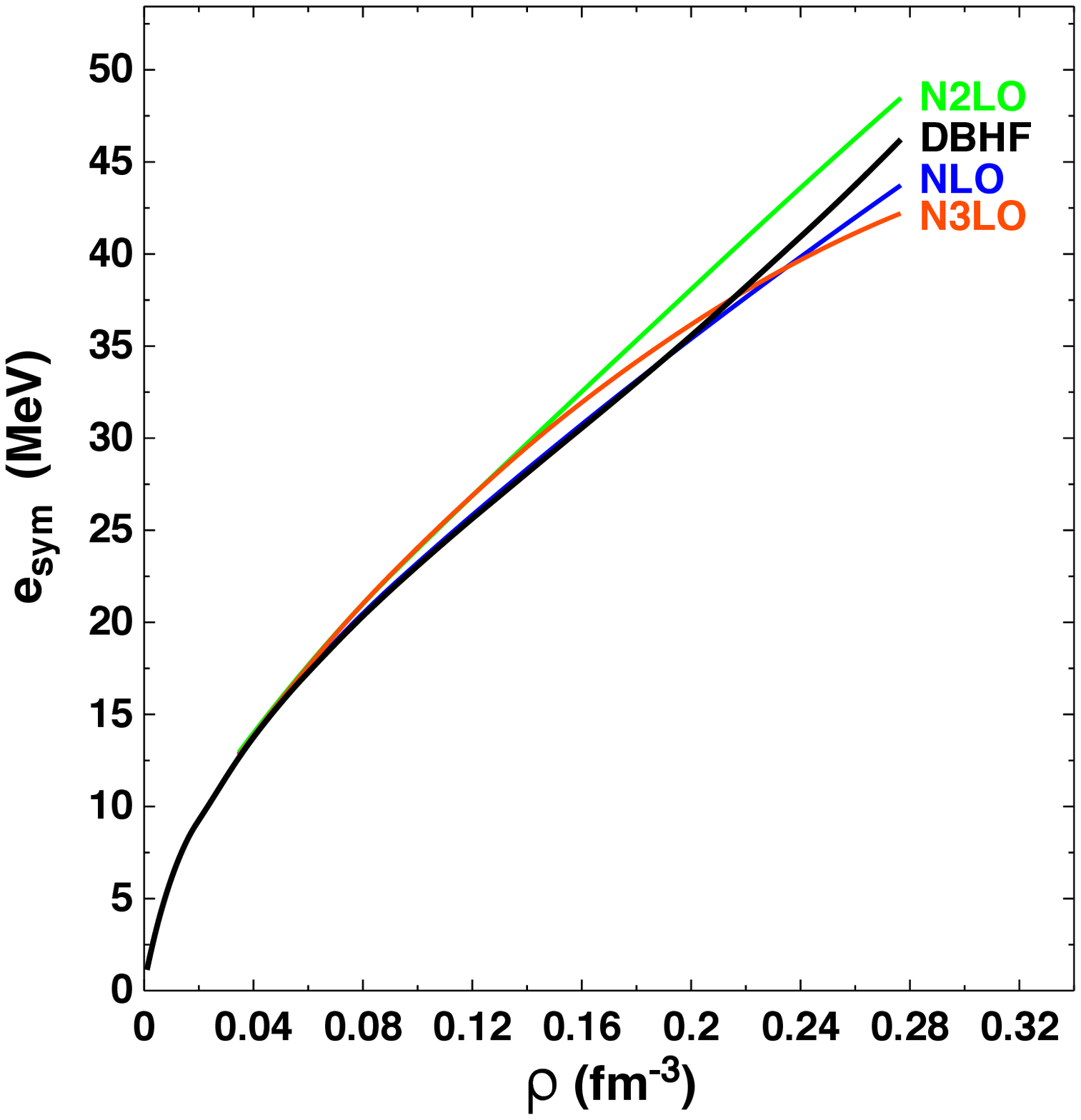}
\vspace*{-0.1cm}
\caption{(Color online) Symmetry energy as a function of density. The DBHF-based predictions 
(black curve) are compared with those from chiral EFT obtained from the EoS shown in Fig.~\ref{snm_nm} 
at the corresponding orders of the chiral expansion. 
} 
\label{esym}
\end{figure}
            
A quick look at Fig.~\ref{ea} and Fig.~\ref{snm_nm} gives a clear indication of the different philosophies between
the ``conventional" (meson-theoretic) approach and the EFT approach, 
where chiral EFT predictions are to be analyzed within the spirit of order-by-order 
convergence and truncation error. 

Consistent application of EFT-based potentials in few- and many-body systems requires 
inclusion of all few- and many-nucleon forces which appear at each order of chiral EFT,                
a task of greater and greater complexity with increasing 
order. In fact, we are not yet at the point where all two-, three-, and four-nucleon forces of order
greater than three have been applied in an $A > $ 3 system. For nuclear matter, a common approximation, which we
apply in our ``N$^3$LO" predictions, is to include the leading (N$^2$LO) 3NF together with the N$^3$LO 2NF.          
On the other hand, thanks to recent progress in the development of chiral NN forces,\cite{n4lo} 
calculations can be conducted in the many-body system with high-precision 2NF                         
up to fifth order. 
Although the predictions thus obtained are incomplete, they are fully
consistent at the 2NF level and can provide valuable information on 
what is missing. More specifically, observing the order-by-order convergence of such 2NF based 
calculations, one can pin down the effect of 3NF with uncertainty quantification. 

With that in mind, we take a closer look at the different equations of states under our consideration.                                       
We begin with a discussion of Fig.~\ref{snm_dbhf} where the DBHF prediction for 
the energy per nucleon in SNM is shown together with
the chiral order-by-order predictions. Clearly, the truncation error at N$^2$LO is dramatically smaller than 
at NLO, indicating good convergence tendencies. Due to the present limitations mentioned in the 
previous paragraph, predictions at N$^4$LO are not available. Furthermore, the calculations at N$^3$LO are not
fully consistent, since the leading (N$^2$LO) 3NF is employed, see comments above. Nevertheless, assuming 
a truncation error at N$^3$LO comparable to the one at N$^2$LO (hopefully a pessimistic estimate), we see that the
DBHF predictions are within the uncertainty of the chiral expansion. 

On the other hand, going back to Fig.~\ref{ea}, we note that those order-by-order
EFT-based predictions are obtained from 2NF only, making it possible to carry the
calculations to N$^4$LO. For that purpose, we employ a recently developed fifth-order chiral
potential.\cite{n4lo} First, we note that the order-by-order pattern bears a clear signature of convergence: The
fifth order correction is considerably smaller that the fourth order one.
Given that the 2NF-based predictions are well converged, and together with the observations made in the 
previous paragraph, the discrepancy with the DBHF-based result could provide
a reasonable indication of the size of the missing chiral 3NF at N$^3$LO. 

A similar discussion can be developed around Fig.~\ref{nm_dbhf} for neutron matter. Again, we see indications that 
the chiral expansion may converge within a range which includes DBHF predictions.
As for the case of symmetric nuclear matter, the right frame shows good convergence of the predictions based 
only on 2NF and can provide an estimate for the missing 3NF.
       
At this point, it is interesting to revisit the diagram in Fig.~\ref{3b}, observed earlier to be both 
a many-body and a relativistic effect. First, we recall that chiral EFT is a low-energy theory and can 
only resolve momenta much lower than the chiral symmetry breaking scale, which can 
be set at about 1 GeV. Therefore, the diagram in Fig.~\ref{3b}, which involves an intermediate 
state with an extra mass of about 2 GeV (the virtual nucleon-antinucleon pair), would be part of what 
chiral EFT describes as short-range or contact 3NF contributions.  
Figures~\ref{ea}, \ref{snm_dbhf} and \ref{nm_dbhf} suggest that the DBHF effective 3NF is an important part           
of the total chiral 3NF.

Next we focus on the symmetry energy. Expanding the energy per particle 
in isospin asymmetric matter with respect to the asymmetry parameter, 
\begin{equation}
\alpha=\frac{\rho_{n}-\rho_{p}}{\rho}\,,\label{alphaalon}
\end{equation}
yields
\begin{equation}
e(\rho, \alpha) = e({\rho},0) + \frac{1}{2} \Big (\frac{\partial ^2 e(\rho,\alpha)}{\partial \alpha ^2}\Big )_{\alpha=0}\alpha ^2 +{\cal O}(\alpha ^4) \; . \label{exp}  
\end{equation}
To a very good degree of approximation, we can write 
\begin{equation}
e(\rho, \alpha) \approx e({\rho},0) + e_{sym}(\rho)\alpha ^2 \; ,   \label{e}                    
\end{equation} 
where $e_{sym}$ is the symmetry energy. 
A typical value for $e_{sym}$ at nuclear matter saturation density is 30 MeV, 
with theoretical predictions spreading approximately between 25 and 35 MeV.
In Fig.~\ref{esym}, we show DBHF predictions for the symmetry energy as a function of
density as well as EFT-based predictions corresponding to the equations of state shown in Fig.~\ref{snm_nm}.
Again, we observe that that the relativistic meson-theoretic predictions are within  
the uncertainty of the chiral perturbation expansion. 

In Ref.\cite{Sam11} we discussed the chief role of the isovector mesons, the pion in particular, in  
building up the symmetry energy and addressed differences with the                 
predictions from mean field approaches, both relativistic and 
non-relativistic.                                                       
In non-microscopic approaches based on quantum hadrodynamics (QHD), such as those originally proposed by 
Walecka and collaborators,\cite{Wal74,Serot1} the dynamical degrees of
freedom are essentially included through coupling of the nucleons to the isoscalar scalar
$\sigma$ and vector $\omega$ mesons. (Note that QHD-I models                                    
of nuclear matter are pionless, but Walecka's QHD-II model does include both $\pi$ and $\rho$.) 
In {\it ab initio} models, mesons are tightly constrained by the free-space data and their parameters 
are never readjusted in the medium. 
Furthermore, the contributions from the various mesons are fully iterated, thus 
giving rise to correlation effects. The corresponding predictions can be dramatically
different than those produced in first-order calculations. As we have shown in previous work,\cite{Sam11}
the symmetry energy can be easily understood in terms of the contributions of each meson
to the appropriate component of the nuclear force and the isospin dependence
naturally generated by isovector mesons. Here, we reiterate                                           
that the pion is the most important ingredient in any realistic model of the 
nuclear force. From the EFT point of view, the pion plays an 
even more fundamental role, being the Goldstone boson of the broken
chiral symmetry of low-energy QCD. 

The remaining part of this section is devoted to applications of the equations of state discussed so far, the
DBHF one in particular, to selected problems of contemporary interest. They are: neutron matter pressure 
and neutron skins; polarization properties of asymmetric matter; short-range correlations (SRC) in nuclear 
matter.

\begin{table}                
\centering
\begin{tabular}{|c||c|c|c|}
\hline
Nucleus & Model & $B/A$(MeV) & $S$ (fm) \\ 
\hline     
\hline
$^{208}$Pb & DBHF  & 7.75    & 0.159  \\ 
           &  BHF  & 7.78    & 0.141  \\ 
           &  NLO  & 7.83    & 0.118  \\ 
           &  N$^2$LO  & 7.65    & 0.197  \\ 
           &  N$^3$LO  & 7.71    & 0.172  \\ 
           &  N$^2$LO(2NF)  & 7.73    & 0.163  \\ 
           &  N$^3$LO(2NF)  & 7.80    & 0.131  \\ 
           &  N$^4$LO(2NF)  & 7.81    & 0.134  \\ 
\hline
$^{48}$Ca & DBHF  & 8.48    & 0.168  \\ 
           &  BHF  & 8.49    & 0.159  \\ 
           &  NLO  & 8.51    & 0.146  \\ 
           &  N$^2$LO  & 8.435   & 0.189  \\ 
           &  N$^3$LO  & 8.46    & 0.176  \\ 
           &  N$^2$LO(2NF)  & 8.47    & 0.171  \\ 
           &  N$^3$LO(2NF)  & 8.50    & 0.154  \\ 
           &  N$^4$LO(2NF)  & 8.50    & 0.154  \\ 
\hline
$^{25}$O & DBHF  & 6.63    & 0.488  \\ 
           &  BHF  & 6.66    & 0.474  \\ 
           &  NLO  & 6.71    & 0.447  \\ 
           &  N$^2$LO  & 6.52    & 0.533  \\ 
           &  N$^3$LO  & 6.58    & 0.507  \\ 
           &  N$^2$LO(2NF)  & 6.60    & 0.496  \\ 
           &  N$^3$LO(2NF)  & 6.67    & 0.464  \\ 
           &  N$^4$LO(2NF)  & 6.70    & 0.460  \\ 
\hline
$^{40}$Mg & DBHF  & 6.69    & 0.572  \\ 
           &  BHF  & 6.73    & 0.549  \\ 
           &  NLO  & 6.82    & 0.509  \\ 
           &  N$^2$LO  & 6.53    & 0.641  \\ 
           &  N$^3$LO  & 6.61    & 0.600  \\ 
           &  N$^2$LO(2NF)  & 6.65    & 0.584  \\ 
           &  N$^3$LO(2NF)  & 6.76    & 0.534  \\ 
           &  N$^4$LO(2NF)  & 6.79    & 0.529  \\ 
\hline
\end{tabular}
\caption                                                    
{ Binding energy per nucleon ($B/A$)                                                  
and neutron skin ($S$) for various neutron-rich nuclei. For each nucleus, the predictions are obtained
with the EoS of NM as indicated in the second column. 
} 
\label{tab1}
\end{table}

\subsection{Neutron skins and the pressure in neutron matter\label{sect31}} 
\label{skins} 

Intense experimental effort is going on to obtain reliable empirical information on the EoS of 
neutron-rich matter.\cite{Tsang+}  Heavy-ion reactions, for instance, are a typical ``tool" to seek constraints 
on the symmetry energy through analyses of observables that are sensitive to the difference between 
the pressure in nuclear and neutron matter, such as isospin diffusion data.                                                                                

The neutron skin of neutron-rich nuclei is sensitive to the   
density derivative of the symmetry energy, which determines to which extent neutrons move 
outwards to form the skin. Parity-violating electron scattering experiments                                     
at low momentum transfer are especially suitable to probe neutron densities,        
due to the dominant coupling of the $Z^0$ boson to the neutron.                
From the first electroweak observation of the neutron skin in a neutron-rich heavy nucleus, a values of           
0.33$^{+0.16}_{-0.18}$ fm for the neutron skin of $^{208}$Pb was determined,\cite{Jlab} but                                                      
the next PREX experiment aims at measuring the skin within an uncertainty smaller by a factor
of 3 (see Ref.\cite{Jlab} and references therein). Potentially, the neutron skin 
of $^{48}$Ca will also be extracted at the Jefferson Laboratory (``CREX" experiment).\cite{crex}
 
The location of neutron drip lines is another issue of great contemporary 
interest which is closely related to the nature of the EoS for neutron-rich matter.       
If a nucleus is extremely neutron-rich, nuclear binding may become insufficient to hold it 
together and the neutron separation energy, defined as $S_n = B(Z,N)-B(Z,N-1)$, where $B$ is the binding 
energy, can be negative, indicating that the last neutron has become
unbound. At this time, the neutron drip line is experimentally accessible only for light nuclei. However,
thanks to the development of radioactive beam facilities, in the near future it may become possible
to explore the stability lines of nuclei ranging from light to very heavy.                       
Note, also, that nuclei beyond the neutron drip lines can exist in the crust of
neutron stars. Those nuclei are believed to determine, for instance, the dynamics of superfluid neutron
vortices, which, in turn, control the rotational properties of the star. 
Very neutron-rich isotopes of Oxygen, Magnesium, and Aluminum have been found. 
For the Oxygen isotopic chain, currently $^{25}$O and $^{26}$O are at the limit of experimental 
availability.\cite{oxy} With regard to Magnesium and Aluminum, $^{40}$Mg and $^{42}$Al are predicted
to be drip line nuclei,\cite{Bau+} suggesting that the drip lines may be located towards heavier isotopes
in this region of the nuclear chart. 

In short, understanding the properties of nuclei with extreme neutron-to-proton ratios is an important and
challenging problem for both rare isotope beam experiments and theoretical models. 
With regard to the theoretical side, {\it microscopic} calculations are important to guide on-going and planned 
measurements.                                                        
                    
We calculate proton and neutron density distributions with a method described in an earlier work.\cite{skin15}   
The method is based on an energy functional derived from the semi-empirical mass formula, where the volume and  
symmetry terms are contained in the isospin-asymmetric equation of state. Thus, we write the 
energy of a (spherical) nucleus as 
\begin{displaymath} 
E(Z,A) = \int d^3 r~ e(\rho,\alpha)\rho(r) + 
\int d^3 r f_0(|\nabla \rho|^2 + \beta 
|\nabla \rho_I|^2) +                 
\end{displaymath} 
\begin{equation} 
\; \; \; \; \; \; \; \; \; \; +\frac{e^2}{4 \pi \epsilon_0}(4 \pi)^2 
\int _0^{\infty} dr' r' \rho_p(r')       
\int _0^{r'} dr r^2 \rho_p(r) \; ,  
\label{drop} 
\end{equation} 
with $\rho(r)$ and $rho_I(r)$ representing  the isoscalar $(\rho_n + \rho_p)$ and isovector density  $(\rho_n - \rho_p)$, respectively.
In the above equation, 
$e(\rho,\alpha)$ is the energy per particle in 
isospin-asymmetric nuclear matter defined in Eq.~(\ref{e}). 
We take the constant $f_0$          
in Eq.~(1) equal to 66 MeV fm$^5$,                  
consistent with Ref.\cite{Oya2010}, 
 whereas  the magnitude of  $\beta$ is about    
1/4.\cite{Furn}  
(Even with variations of $\beta$ between -1 and +1, we find that     
the size of that term is very small, so we ignore its 
contribution.) 

The proton and neutron density functions are obtained by minimizing the value
of the energy, Eq.~(\ref{drop}), with respect to the parameters of Thomas-Fermi distributions
for proton and neutron densities.  This method has the advantage of allowing for a         
very direct connection between the EoS and some bulk properties of finite nuclei, which is the point of our study.    
Clearly, our method does not account for shell or pairing effects, as our 
purpose is not to perform detailed structure calculations, but rather to highlight the 
direct impact of the equation of state on the nuclear properties under consideration. 

Here, we wish to emphasize the role of neutron matter pressure on the neutron skin thickness. Clearly, not all
the EoS discussed above are realistic, particularly with regard to SNM saturation.
To avoid excessively unrealistic values of the binding energy, in what follows we take the 
EoS of SNM from phenomenology\cite{snm} and keep the focus on the (microscopically 
predicted) NM pressure and its impact on the neutron skin. 

\begin{figure}
\begin{center}
\vspace*{-1.0cm}
\includegraphics[width=6.0cm]{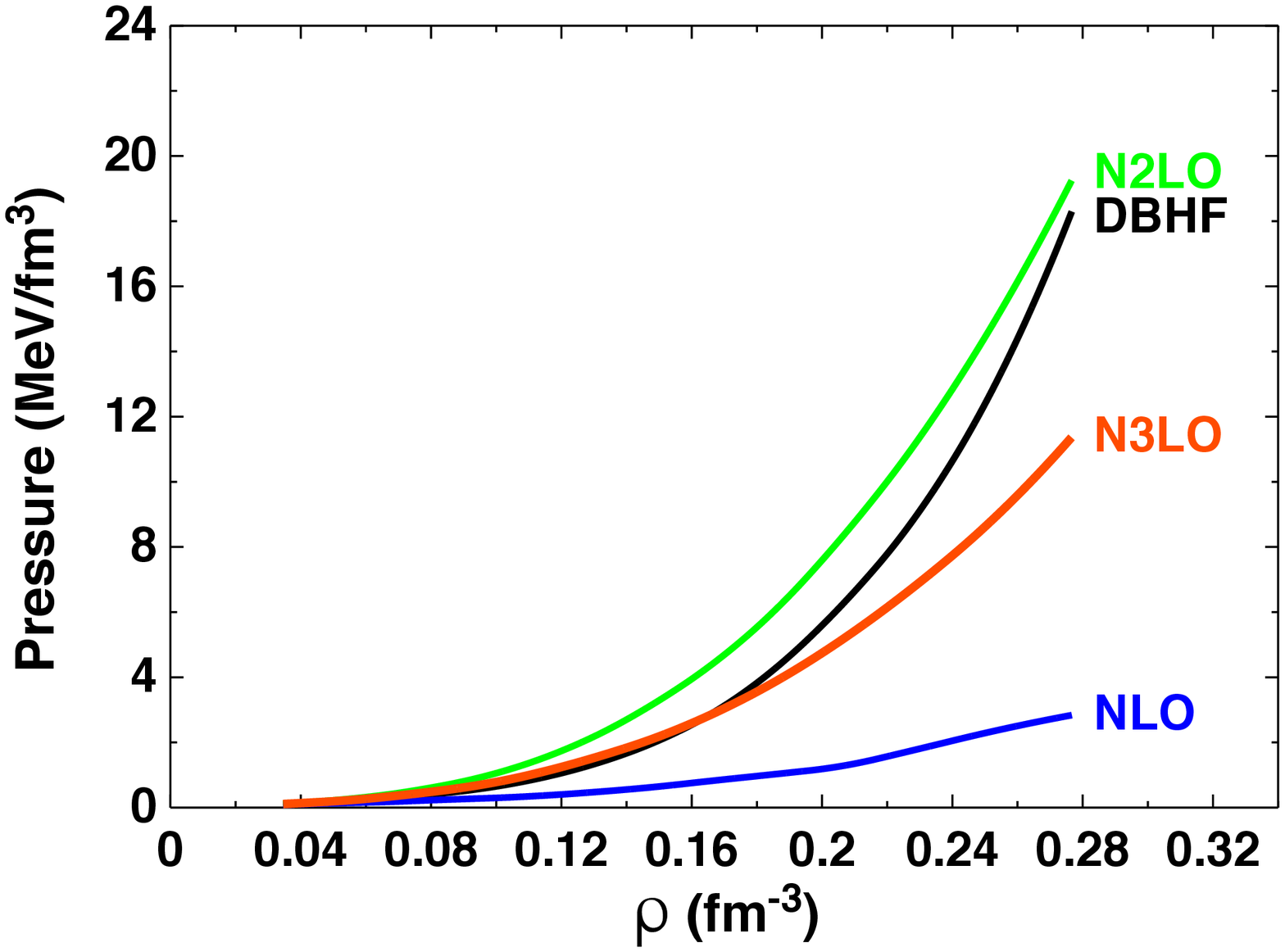}
\includegraphics[width=6.0cm]{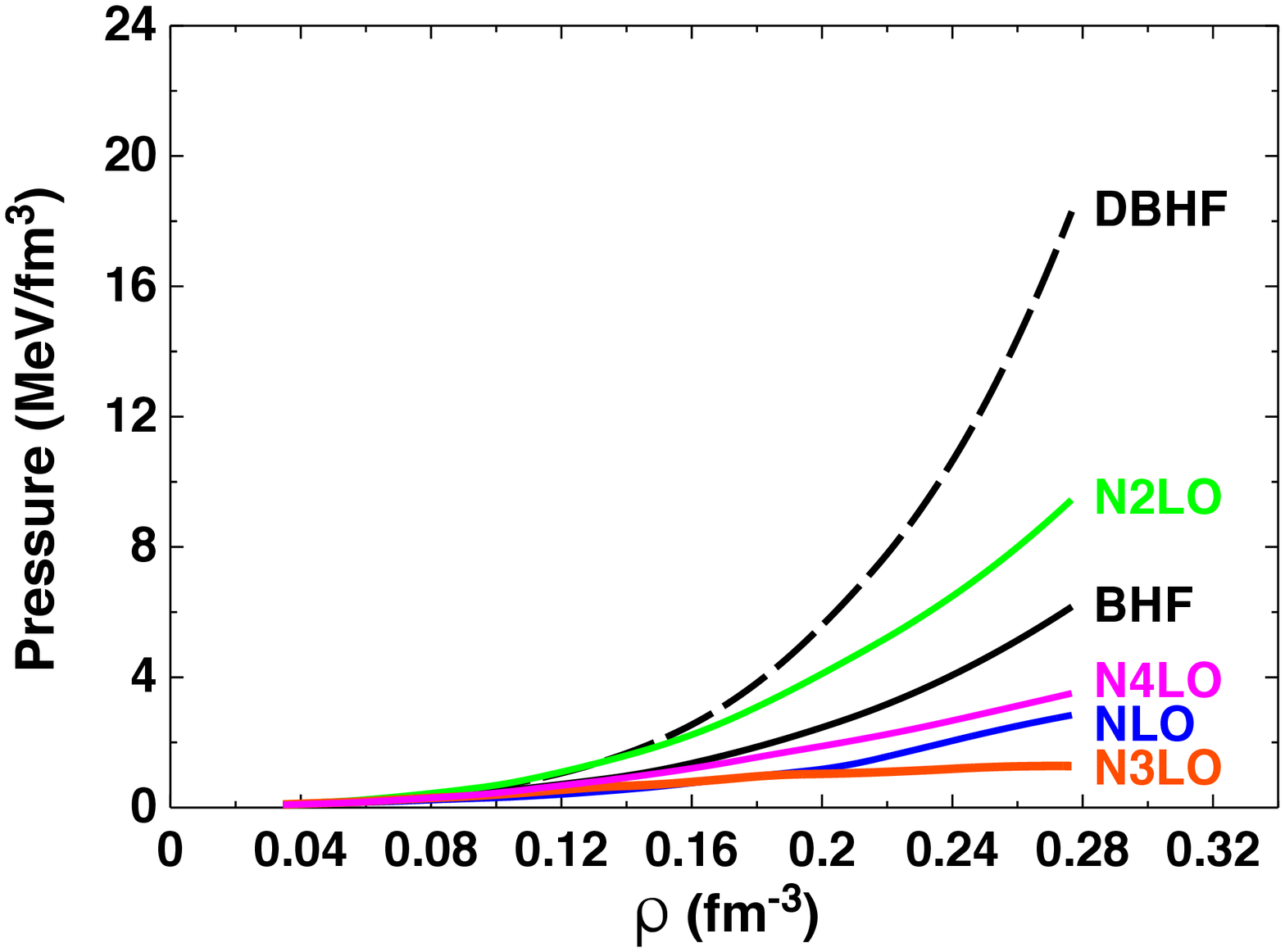}
\vspace*{-0.1cm}
\caption{Left: Pressure in neutron matter for the EoS displayed in the left frame of Fig.~\ref{nm_dbhf}. Right:   
 Pressure in NM for the EoS displayed in the right of Fig.~\ref{nm_dbhf}.                 
} 
\label{press}
\end{center}
\end{figure}

On the left and right sides of Fig.~\ref{press}, we show the pressure in neutron matter for the different models 
included in the left and right panels of Fig.~\ref{nm_dbhf}, respectively.
The corresponding neutron skin predictions for $^{208}$Pb and $^{48}$Ca as well as two 
very neutron-rich (presumed drip) nuclei are found in Table~\ref{tab1}. 
Note that the {\it average} value of the asymmetry parameter $\alpha$, which can be defined as 
$(N-Z)/A$, is equal to 0.36 and 0.40 for $^{25}$O and $^{40}$Mg, respectively, much larger than the $^{208}$Pb value 
of 0.21. Hence, the much larger skin seen in these nuclei. 

As expected, larger neutron matter pressure implies larger skin, although the variations in the skin are not 
as dramatic as one might expect when looking at the NM pressure. This is because the neutron asymmetry parameter
 as a function of the radial coordinate, $\alpha(r)$, becomes appreciably greater than
zero only in the peripheral area of the nucleus, namely where the density is much lower than normal (central)
densities. Therefore, if different models have similar pressure values at low density, even though they differ
much more strongly at higher density, the typical pressure-skin correlation may not be clearly identifiable. 

In $^{48}$Ca and $^{208}$Pb the DBHF and the N$^3$LO predictions differ by about 0.01 fm. As we argued earlier, 
the well-converged results at N$^4$LO with only 2NF can be exploited to estimate the uncertainty in PREX and 
CREX experiments which would allow to discern the impact of 3Nf on the skin.\cite{FS16}

\subsection{Polarization in asymmetric matter\label{sect32}} 
\label{pol} 
                                                                     
In this subsection we address the issue of polarization in isospin asymmetric matter.
Such system has gathered much attention, especially in conjunction with the possibility of ferromagnetic instabilities  
in the interior of pulsars. The presence of polarization would impact, for instance, neutrino cross sections and 
luminosities and thus the mechanism of neutron star cooling. At this time, conclusions with regard 
to phase transitions into spin ordered states are often contradictory, both quantitatively and qualitatively. 

In Ref.\cite{Sam11spin} we extended our DBHF calculation to matter 
with arbitrary degree of spin and isospin asymmetry. Our findings did not show evidence of 
a phase transition to a ferromagnetic (FM) or antiferromagnetic (AFM) state. We observed that such                 
conclusion appears to be shared by predictions of microscopic models, such as those based on conventional
Brueckner-Hartree-Fock theory.\cite{VB02} 
On the other hand, calculations based on effective forces suggest different conclusions. For instance, 
with the {\it SLy4} and {\it SLy5} Skyrme forces and the Fermi liquid formalism,                                       
a phase transition  to the AFM state is predicted in asymmetric matter 
at a critical density equal to about 2-3 times normal density.\cite{IY}  
Qualitative disagreement is also encountered with other non-microscopic approaches such as                     
relativistic Hartree-Fock models based on effective meson-nucleon Lagrangians. For instance, in Ref.\cite{Marc91} 
it was reported that the onset of a ferromagnetic transition in neutron matter, and its critical density, 
are sensitive to the inclusion of isovector mesons and the nature of their couplings. 

Polarized NM is a very interesting system for other reasons as well, which do not involve
the  very high densities found in the core of compact stars. 
Because of the large neutron-neutron scattering length, NM displays behaviours similar to those 
of a unitary Fermi gas. In fact, up to nearly normal density, unpolarized NM was found to behave like an                  
$S$-wave superfluid\cite.{Carls03,Carls12} The possibility of simulating low-density NM with 
ultracold atoms near a Feshbach resonance\cite{Bloch08} has also been discussed.
When the system is totally polarized, it has been observed to resemble a weakly interacting Fermi gas.\cite{krueg14}  

More recently,\cite{SMK15} these issues have been revisited in a first calculation including 
both spin and isospin asymmetries within the framework of chiral forces.
In-medium effective chiral 3NF are derived which are suitable for the most general case
of different proton and neutron concentrations where each species can be polarized to a different degree.

The formalism we use to obtain the polarized matter EoS within the DBHF 
framework was explained in detail in Refs.\cite{Sam11spin} and reviewed in Ref.\cite{ijmpe13}, whereas the details of
the recent chiral calculations can be found in Ref.\cite{SMK15}. Here, consistent with the spirit of the 
section, we will look at DBHF and EFT predictions side by side and draw conclusions as appropriate.                          

The following definitions will be useful.
In a spin polarized and isospin asymmetric system with fixed total density, $\rho$,               
the partial densities of each species are               
\begin{equation}
\rho_n=\rho_{nu}+\rho_{nd}\; , \; \; \; 
\rho_p=\rho_{pu}+\rho_{pd}\;, \; \; \; 
\rho=\rho_{n}+\rho_{p} \; ,           
\label{rho} 
\end{equation}
where $u$ and $d$ refer to up and down spin polarizations, respectively, of protons ($p$) or neutrons ($n$). 
The isospin and spin asymmetries, $\alpha$, $\beta_n$, and $\beta_p$,  are defined in a natural way: 
\begin{equation}
\alpha=\frac{\rho_{n}-\rho_{p}}{\rho} \;, \; \; \;
\beta_n=\frac{\rho_{nu}-\rho_{nd}}{\rho_n} \;, \; \; \; 
\beta_p=\frac{\rho_{pu}-\rho_{pd}}{\rho_p} \;. 
\label{alpbet} 
\end{equation}
The density of each individual component can be related to the total density by 
\begin{displaymath}
\rho_{nu}=(1 + \beta_n)(1 + \alpha){\rho \over 4} \;,           
\rho_{nd}=(1 - \beta_n)(1 + \alpha){\rho \over 4}\; ,           
\end{displaymath}
\begin{equation}
\rho_{pu}=(1 + \beta_p)(1 - \alpha){\rho \over 4}\; ,          
\rho_{pd}=(1 - \beta_p)(1 - \alpha){\rho \over 4}\; , 
\label{rhopnud}
\end{equation}
where each partial density is related to the corresponding Fermi momentum 
through $\rho_{\tau \sigma}$ =$ (k_F^{\tau \sigma})^3/(6\pi^2)$, $\tau=p,n$ and $\sigma=u,d$. 
The {\it average} Fermi momentum  and the total density are related in the usual way as 
$\rho= (2 k_F^3)/(3 \pi ^2)$. 

\begin{figure}[!htb] 
\centering 
\hspace*{-1.0cm}
\scalebox{0.50}{\includegraphics{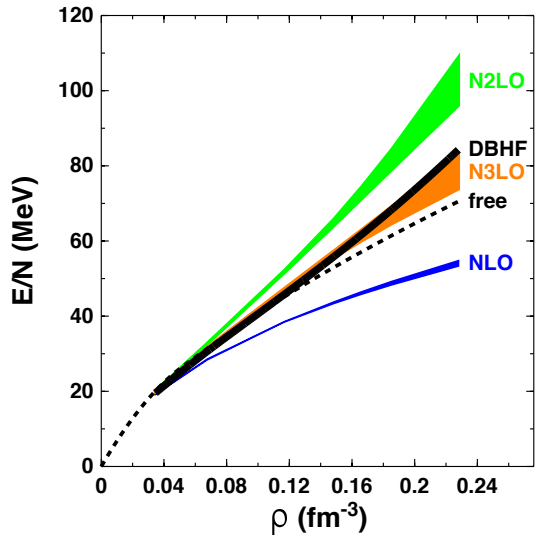}} 
\vspace*{-0.1cm}
\caption{(color online)                                        
DBHF predictions for the 
energy per neutron in fully polarized neutron matter as a function of density (solid black). 
EFT-based results are also shown at different orders of the chiral expansion (notation as in Fig.~\ref{snm_nm} ). At each order, the bands
are obtained from varying the cutoff in the regulator function between 450 and 600 MeV.               
The dotted curve shows the energy of the free Fermi gas.
} 
\label{pnm}
\end{figure}

In Fig.~\ref{pnm}, the solid black curve is the DBHF prediction for the energy per neutron in fully polarized matter
as a function of density. The blue, green, and orange bands represent the predictions of 
the EFT-based calculations, with the notation specifying the chiral order the same as in the previous sections. 
The width of each band is obtained by changing the cutoff in the regulator function of the chiral 
potentials\cite{obo,SMK15} between 450 MeV and 600 MeV, so as to explore to which extent conclusions 
may depend on the resolution scale. Cutoff dependence is very weak up to saturation density and
generally moderate. Some indication of slow convergence can be seen when moving from 
N$^2$LO to N$^3$LO calculation. 

The predictions from DBHF and N$^3$LO are close to each other and to the free Fermi gas energy 
(dashed black line), at least up to saturation densities.                                                                                             
With regard to the similarity with the free Fermi gas, it is interesting to include some additional considerations.
As mentioned earlier, many-fermion systems with large scattering lengths offer the 
opportunity to model low-density neutron matter. In the unitary limit (that is, when the system can support a 
bound state at zero energy), the scattering length approaches infinity. The system then becomes
scale-independent and the ground state energy is determined by a single universal parameter, known 
as the Bertsch parameter, $\xi$. The latter is defined as the ratio of the energy per particle of the unitary gas
to that of the free Fermi gas. In Ref.~\cite{Kais12}, using a simple ansatz for the interaction,
it is shown that $\xi$ increases from approximately 0.5 to 1.0 as the spin asymmetry of neutron matter is 
increased from 0 (unpolarized) to 1 (fully polarized).

\begin{figure}[!htb] 
\centering 
\vspace*{-0.5cm}
\hspace*{-1.0cm}
\scalebox{0.50}{\includegraphics{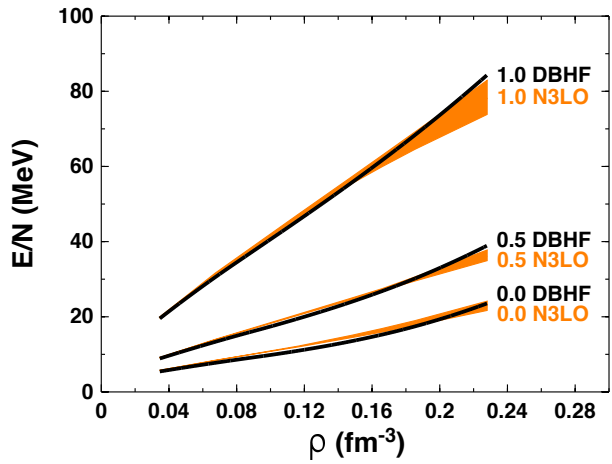}} 
\vspace*{-0.1cm}
\caption{(color online)                                        
Black solid curves:
Energy per neutron in pure neutron matter as a function of density as 
predicted with DBHF. From lowest to highest curve:
unpolarized NM; partially polarized NM, with $\beta_n$=0.5; fully polarized NM ($\beta_n$=1). 
The three orange bands show EFT predictions at N$^3$LO under the same conditions. 
The width of each band shows the uncertainty from varying the cutoff between 
450 and 600 MeV.                                                                                       
} 
\label{beta}
\end{figure}

In Fig.~\ref{beta}, we compare DBHF predictions (solid black lines) of the energy per neutron 
with those from the N$^3$LO calculations along with their cutoff variations 
(orange bands) in: unpolarized NM (lowest curve and band); partially polarized NM ( curve and 
band in the middle); and fully polarized NM (highest curve and band). 
For the partially polarized case, the value of the neutron spin asymmetry parameter, $\beta_n$, is equal to 0.5, 
corresponding to 75\% of the neutrons being polarized in one direction and 25\% in the opposite direction, 
see Eq.~(\ref{rhopnud}). Clearly, a lesser degree of spin asymmetry (as compared to the ferromagnetic case) yields 
considerably less repulsion. 
There is definitely no sign of a phase transition, particularly to a ferromagnetic state, nor an indication that 
such transition may occur at higher densities. These observations apply to both DBHF and 
N$^3$LO, whose predictions are clearly very close. 

\begin{figure}[!htb] 
\centering 
\hspace*{-1.0cm}
\scalebox{0.50}{\includegraphics{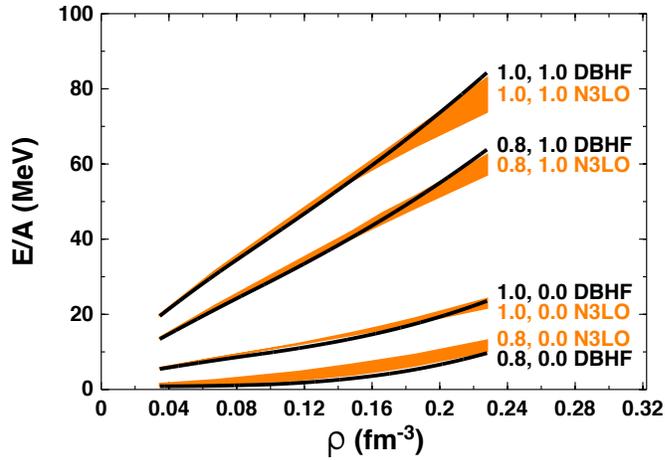}} 
\vspace*{-0.1cm}
\caption{(color online)                                        
Solid black lines: DBHF predictions for the energy per nucleon in neutron-rich matter 
as a function of density at different conditions of isospin and spin polarization.
The curve labeled as ``0.8, 1.0" displays the results for neutron-rich matter with a 
proton fraction equal to 10\% ($\alpha$=0.8) and fully polarized neutron ($\beta_n$=1.0).                      
The curve labeled as ``0.8, 0.0" refers to neutron-rich matter with the same proton fraction 
and no polarization ($\beta_n$=0.0). The protons are unpolarized. 
The orange bands are the corresponding EFT predictions.         
As a reference point, we also include the DBHF curves and EFT bands already shown in the previous figure, 
which refer to pure neutron matter ($\alpha$=1) with fully polarized ($\beta_n$=1) or 
unpolarized ($\beta_n$=0) neutrons. The bands are obtained varying the cutoff between 450 and 600 MeV.
} 
\label{alpha}
\end{figure}

Most typically, models which do predict spin instability of neutron matter find the phase transition to 
occur at densities a few times normal density. Such high densities are outside the domain of chiral perturbation
theory. 
With some effective forces, though, it was found\cite{pol19} that a small fraction of protons can significantly 
reduce the onset of the threshold density for a phase transition to a spin polarized state of neutron-rich matter.
We explored this scenario by adding a small fraction of protons to fully polarized or unpolarized neutrons.
From Eqs.~(\ref{rho}-\ref{rhopnud}), a proton fraction of 10\% is obtained with $\alpha$=0.8.
The results are displayed in Fig.~\ref{alpha}, where a crossing of the solid black curves (or bands)
labeled as ``0.8, 1.0" and ``0.8, 0.0", respectively, would indicate a phase transition. Thus we conclude that such
transition is not predicted by the relativistic meson-theoretic model or with chiral forces. By extrapolation,
a transition to a polarized state would also appear very unlikely at higher densities.

In summary, in both the DBHF and the N$^3$LO calculations (regardless the cutoff),                 
the energies of the unpolarized system at normal density are 10-20 MeV,          
whereas those in the polarized case are above 60 MeV. Thus, even allowing for a large 
uncertainty (see comments above), a phase transition to a ferromagnetic state can be excluded.
This conclusion remains valid in the presence of a small proton fraction.

\subsection{Momentum distribution and short-range correlations in nuclear matter\label{sect33}} 
\label{src} 

In this section, we discuss correlations in nuclear matter, short-range and tensor correlations in particular, from
the DBHF perspective. 

Since the early Brueckner nuclear matter calculations,\cite{HT70} it has been customary to associate the 
correlated two-body wave functions to the strength of the $NN$ potential. When this is done 
in a particular channel, one can extract information about specific components of the force.
For instance, the $^3S_1-^3D_1$ channel reveals information on tensor correlations, which 
have traditionally attracted particular attention, since the model dependence among predictions from different 
$NN$ potentials resides mostly in the strength of their respective tensor forces and off-shell behaviors.\cite{polls2000} 
                                                                  
This topic has been addressed numerous times in the literature and has recently attracted 
renewed attention in conjunction with empirical analyses of electron scattering measurements at high
momentum transfer. We will address those at the end of this section. 
In short, investigations of short-range correlations are of 
contemporary interest and appropriate in a discussion of microscopic in-medium interactions. 
Other recent studies of tensor correlations can be found in Refs.\cite{Carb13,Rios14}, where 
the self-consistent Green's function method is used to obtain single-particle properties.  
In Ref.\cite{Wir+} both single-nucleon and nucleon-pair momentum distributions in 
$A \leq$ 12 nuclei are addressed.                                                      

First we review some basic concepts leading to the definition of the defect function and the 
wound integral, both closely related to the correlated wave function.                                     
                                                                     
In terms of relative and center-of-mass momenta, the Bethe-Goldstone equation, Eq.~(\ref{eq:bg}), can be written as 
\begin{equation}
G({\bf k}_0, {\bf k},{\bf P}^{c.m.}, E_0) = V({\bf k}_0, {\bf k}) +\int d^3{\bf k}^{'} \,V({\bf k}_0, {\bf k}^{'})        
\frac{Q(k_F,{\bf k}^{'}, {\bf P}^{c.m.})}{E_0-E}           
G({\bf k}^{'}, {\bf k},{\bf P}^{c.m.}, E_0) \; ,                                                                         
\label{BG} 
\end{equation}
where $V$ is the $NN$ potential, $Q$ is the Pauli operator, $E = {\cal H}({\bf k}^{'}, {\bf P}^{c.m.})$, and 
$E_0 = {\cal H}({\bf k_0}, {\bf P}^{c.m.})$, with the function ${\cal H}$ the total energy of the two-nucleon pair. 

The second term on the right-hand side of Eq.~(\ref{BG}) represents the infinite ladder sum which contains the effects of correlations
in the wave function. The correlated ($\psi$) and the uncorrelated ($\phi$) two-particle
wave functions are related through 
\begin{equation}
G\phi = V \psi  \; , 
\label{wf} 
\end{equation}
which implies 
\begin{equation}
\psi = \phi + V \frac{Q}{E_0-E}G\phi \; . 
\label{psi} 
\end{equation}
In the two equations above, we switched to operator notation for simplicity.
The difference between the correlated and the uncorrelated wave functions, 
$f=\psi - \phi$, is referred to as the defect function and is clearly a measure of correlations.                        
It is convenient to consider its momentum-dependent Bessel transform, which gives, for each angular 
momentum state, (and average center-of-mass momentum $P_{avg}^{c.m.}(k_0,k_F)$),             
\begin{equation}
f_{LL'}^{JST}(k,k_0,k_F) = \frac{k \; \bar{Q}(k_F,k,P_{avg}^{c.m.}) G_{LL'}^{JST}(P_{avg}^{c.m.},k,k_0)}{E_0-E} \;, 
\label{ff} 
\end{equation}
where the angle-averaged Pauli operator has been employed to restrict correlation effects to partial waves with fixed total angular momentum. 
This is related to the probability of exciting two nucleons with relative momentum $k_0$ and relative orbital
angular momentum $L$ to a state with relative momentum $k$ and relative orbital
angular momentum $L'$.                   
The integral of the probability amplitude squared is known as the wound integral  
and defined, for each partial wave at some density $\rho$, as 
\begin{equation}
\kappa_{LL'}^{JST}(k_0,k_F) = \rho \int_0^{\infty} |f_{LL'}^{JST}(k,k_0,k_F)|^2 dk \; .         
\label{kappa} 
\end{equation}
Thus, $f$ and $\kappa$ provide a measure of correlations present in the wave function and the 
$G$-matrix. 

In the present calculations, we take the initial momentum equal to 0.55$k_F$. This value is the $r.m.s.$ 
value of the relative momentum of two nucleons having an average center-of-mass momentum, 
$P_{avg}^{c.m}$,  such that  their initial momenta in the nuclear matter rest frame, $k_1$ and $k_2$, 
are below the Fermi sea. With these constraints (see Ref.\cite{HT70} )

\begin{figure}[!htb] 
\centering         
\vspace*{-1.2cm}
\hspace*{0.5cm}
\scalebox{0.25}{\includegraphics{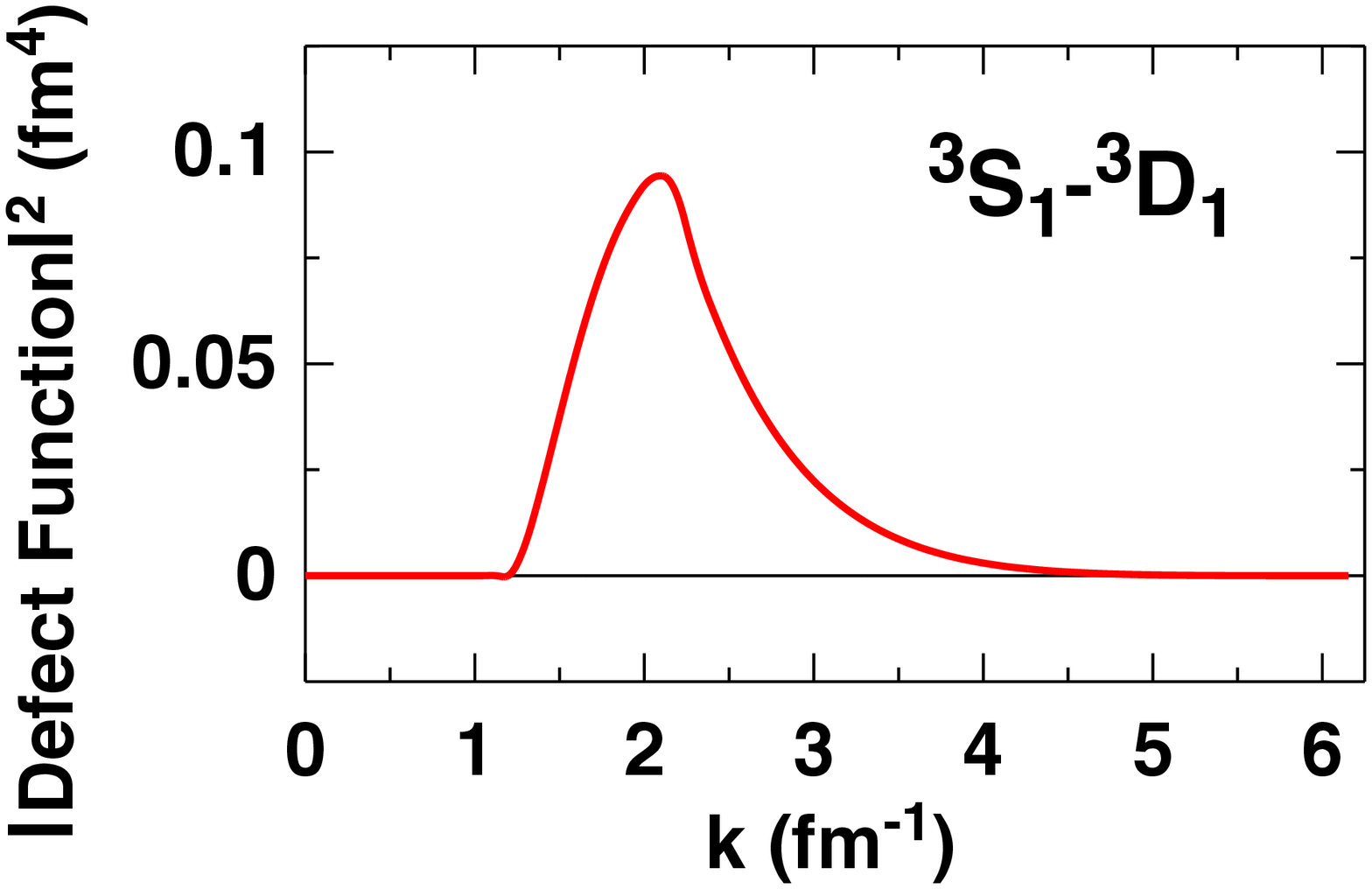}}
\scalebox{0.25}{\includegraphics{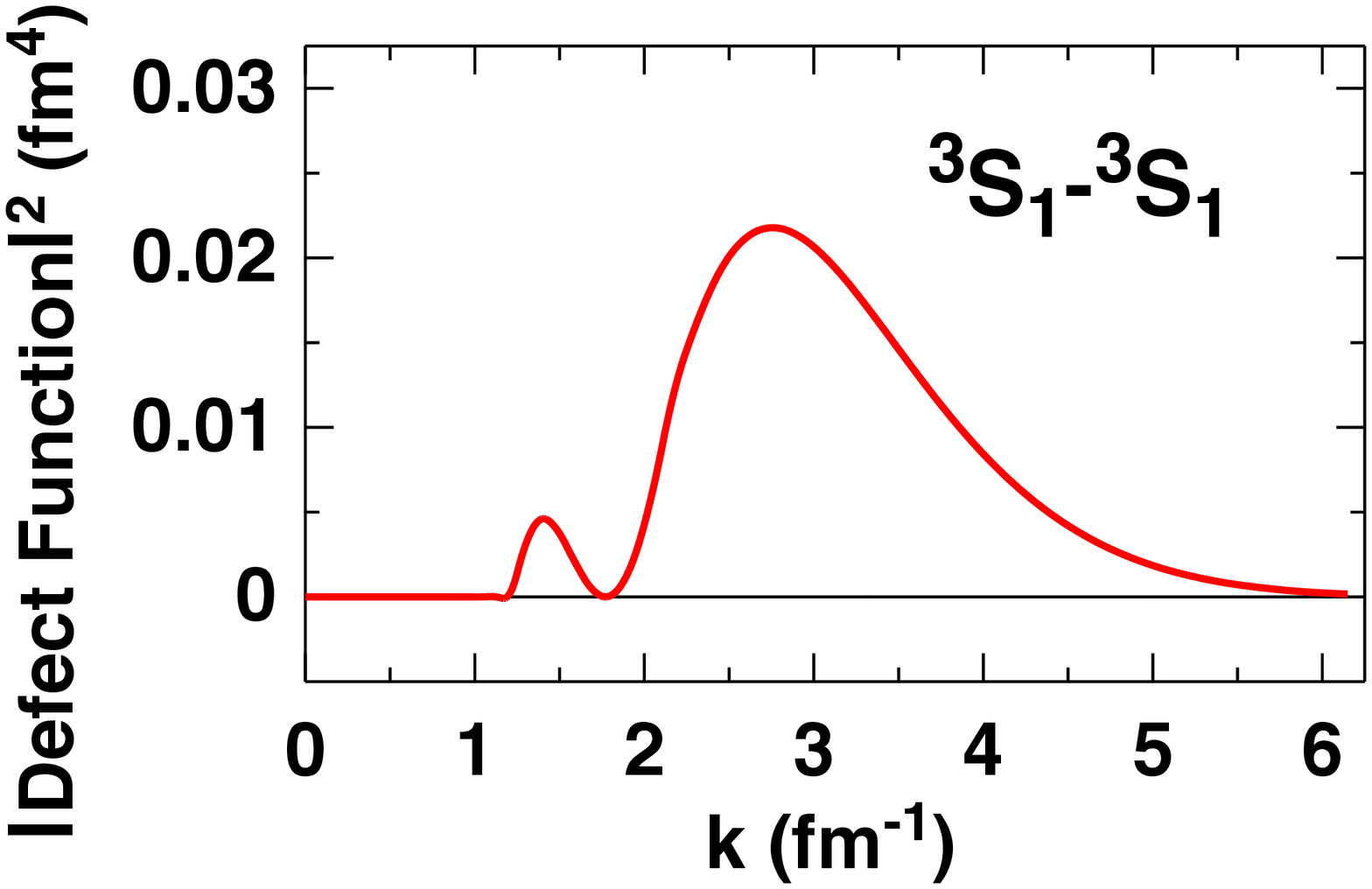}}
\vspace*{-2.2cm}
\caption{(Color online)                                                         
Magnitude squared of the defect function Eq.~(\ref{ff}) for the $^3S_1-^3D_1$ transition (left) and the $^3S_1$ state (right) in symmetric
nuclear matter as predicted with DBHF. The Fermi momentum is equal to 1.4 fm$^{-1}$.                                                                      
}
\label{sd14}
\end{figure}

On the left-hand side of Fig.~\ref{sd14} we show the magnitude squared of the 
defect function, Eq.~(\ref{ff}), for the $^3S_1-^3D_1$ transition as a functions of the final relative
momentum $k$ in symmetric matter. The total density corresponds to a Fermi momentum
of 1.4 fm$^{-1}$, which is close to the saturation density of symmetric nuclear matter. 
Notice that these distributions are excitation probabilities rather than standard 
momentum distributions (which are usually larger at low momenta). In other words, these curves
do not include the distribution of momenta for occupied states below the Fermi surface. 

Clearly there is a high probability that the $np$ pair is excited to a state with relative momentum 
of about 2 fm$^{-1}$ {\it via} a tensor transition. 
For comparison, the same quantity is shown on the right-hand side of Fig.~\ref{sd14} for the $^3S_1$ state.                                      
Note that the latter carries information on short-range central correlations, namely the repulsive core of the
central force, although it is also impacted by the tensor force because of its coupling to the $D$-state. 
It peaks around a momentum of about 3 fm$^{-1}$ and has a distinct node between 1.5 and 2 fm$^{-1}$.                                                                    
Notice that the $^3S_1$ probability amplitude tends to be broader, while the amplitude of the tensor transition 
has a much larger absolute value.

\begin{figure}[!htb] 
\centering         
\vspace*{-1.2cm}
\hspace*{0.5cm}
\scalebox{0.25}{\includegraphics{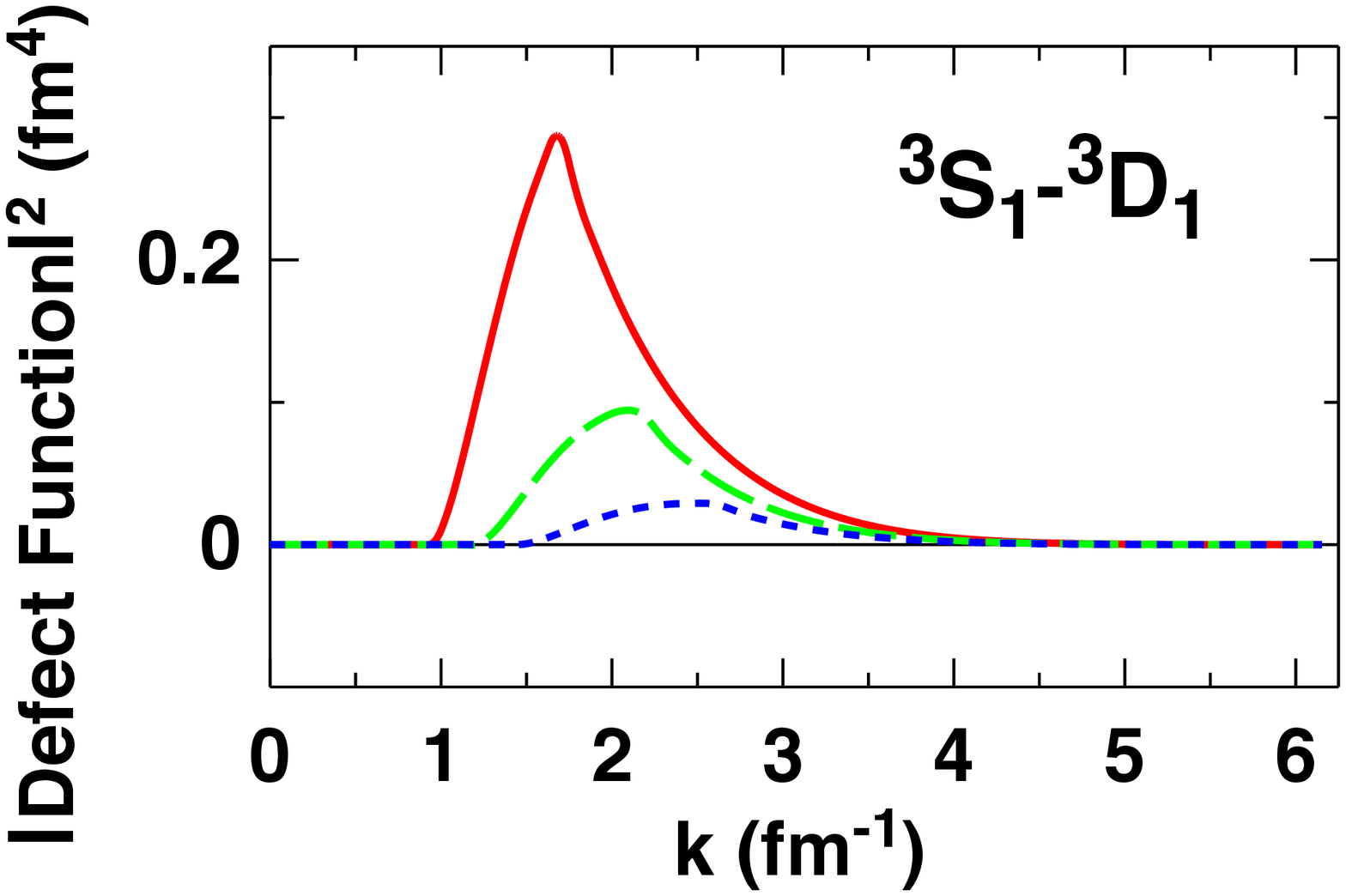}}
\scalebox{0.25}{\includegraphics{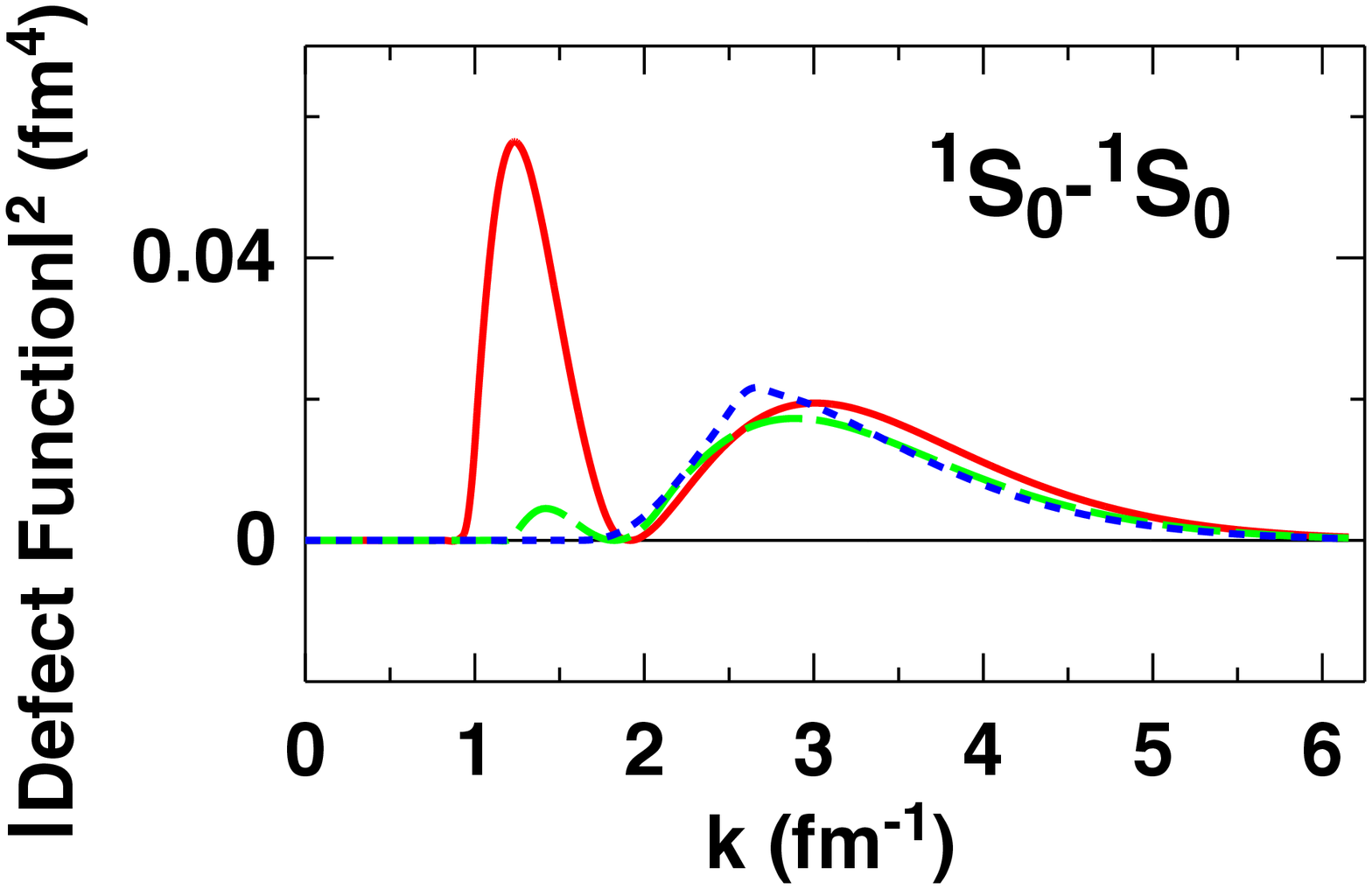}}
\vspace*{-2.8cm}
\caption{(Color online)                                              
DBHF predictions for 
the magnitude squared of Eq.~\ref{ff} at three different values of the Fermi
momentum in symmetric matter: $k_F$=1.1 fm$^{-1}$ (solid red); 
$k_F$=1.4 fm$^{-1}$ (dashed green); 
$k_F$=1.7 fm$^{-1}$ (dotted blue). The left panel shows the effects pf the tensor components in the $^3S_1-^3D_1$ channel,
whereas the $^1S_0$ channel has been considered as an example for the central components in the right panel.
}
\label{all}
\end{figure}

In Fig.~\ref{all} we consider three different densities of symmetric nuclear matter.                    
The defect function for the tensor transition maintains a similar shape with changing density, 
with the peak shifting towards lower(higher) momenta at the lower(higher) density, due to the changing
impact of Pauli blocking in each case.                        
For the $^1S_0$ (as well as for the central part in the $^3S_1$ not displayed here), the peak at the lower momenta grows larger at the lower density.    

The individual contributions to the wound integral, Eq.~(\ref{kappa}), from the states considered in the figures
are shown in Table~\ref{tab1x} for three densities of symmetric matter. The contribution of the central force 
(as seen through $^1S_0$) relative to the tensor force increases with increasing density, due to the enhanced
impact of the repulsive core when higher momenta are probed (as is the case in a system with increasing 
Fermi momentum). 

\begin{table}[!htb]                
\centering \caption                                                    
{Contributions to the wound integral, Eq.~\ref{kappa}, from $J=0$ and $J=1$ states at different densities, as 
predicted with DBHF.
} 
\vspace{5mm}
\begin{tabular}{|c|c|c|c|}
\hline
$ k_F$ (fm$^{-1}$) & $^3S_1-^3D_1$ & $^3S_1-^3S_1$ & $^1S_0$ \\
\hline     
 1.1 &  0.079& 0.025 & 0.017   \\
\hline
 1.4 &  0.060 & 0.022  & 0.019 \\ 
\hline
 1.7 &  0.037 & 0.031 & 0.035 \\ 
\hline
\end{tabular}
\label{tab1x} 
\end{table}

Next, we will compare with the high-precision $NN$ chiral potential at N$^3$LO we already encountered in the previous
sections. We recall that, for chiral interactions, the characteristic momentum scale is below the scale set
by the cutoff in the 
regulator function. For the interaction employed here, the latter has the form 
\begin{equation}
F(p',p) = exp[-(p'/\Lambda)^{2n} - (p/\Lambda)^{2n}] \; ,
\label{reg}
\end{equation}
where $p$ and $p'$  denote, respectively, the magnitudes of the
initial  and  final  momenta of the interacting nuclei  in  the  center-of-mass
frame.  A typical value for $\Lambda$ is 0.5 GeV, which corresponds roughly to 2.5 fm$^{-1}$. Note that this regulator function is
defined in terms of nucleon momenta, while cutoff function in meson theory are expressed in terms of momentum transfer. This
induces a larger non-locality in chiral interactions as compared to those based on meson theory.

Due to the cutoff, chiral potentials are much softer 
than meson-theoretic ones. More precisely, the Bonn B potential vanishes for relative momenta around 
10 fm$^{-1}$,  whereas the chiral $NN$ interaction is essentially
negligible already near 4 fm$^{-1}$.

\begin{figure}[!htb] 
\centering         
\vspace*{-1.2cm}
\hspace*{0.5cm}
\scalebox{0.25}{\includegraphics{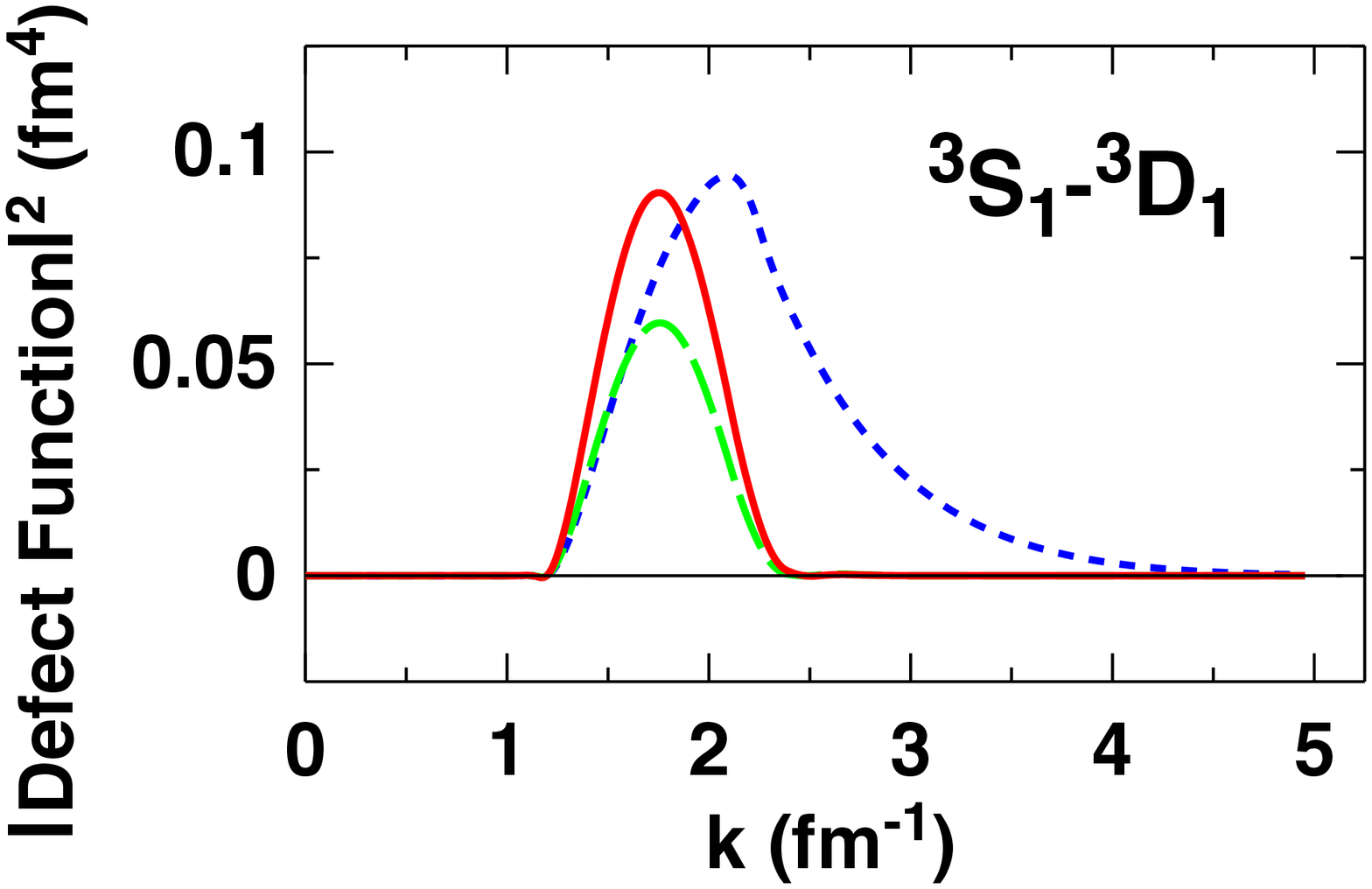}}
\scalebox{0.25}{\includegraphics{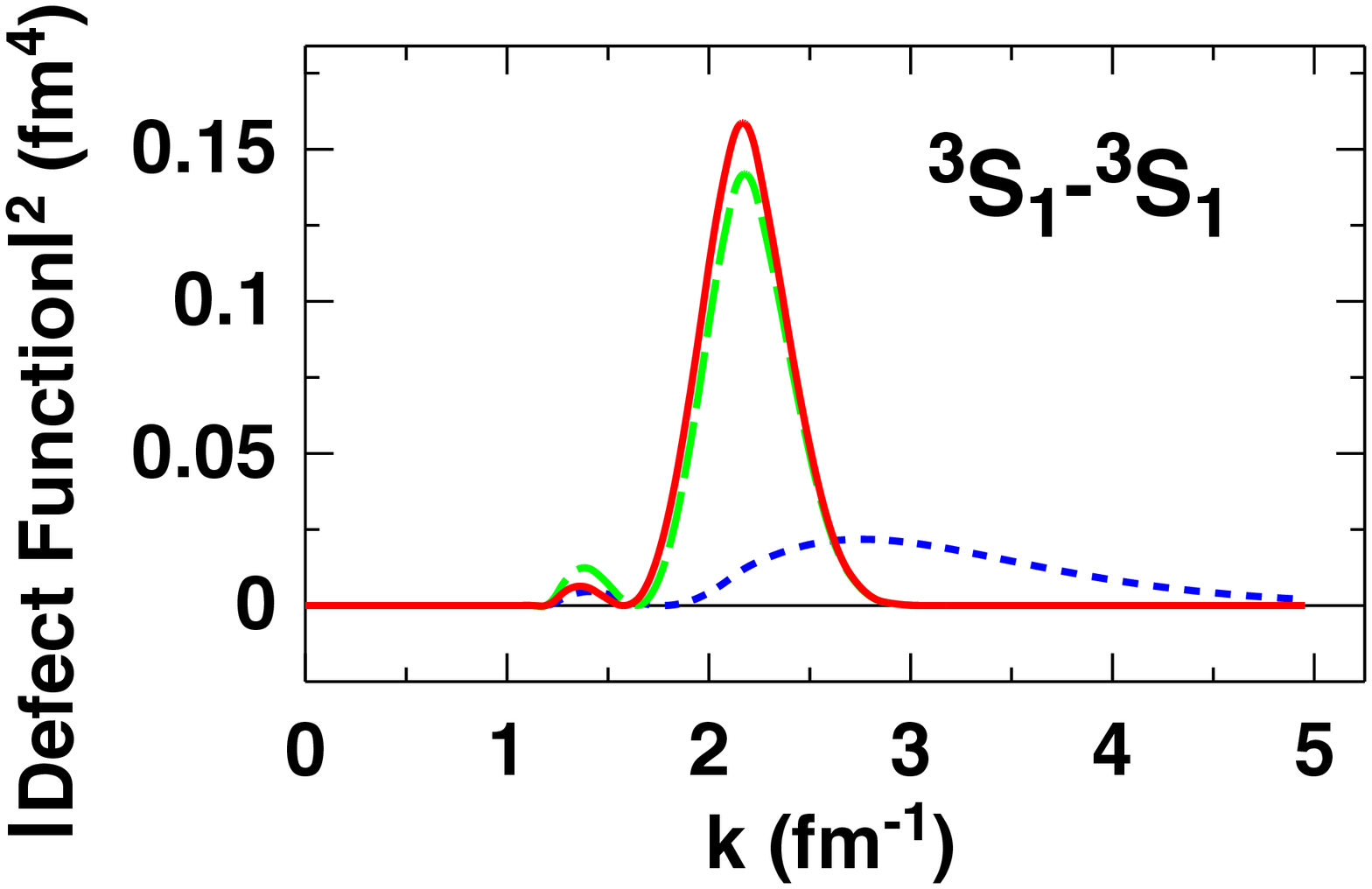}}
\scalebox{0.25}{\includegraphics{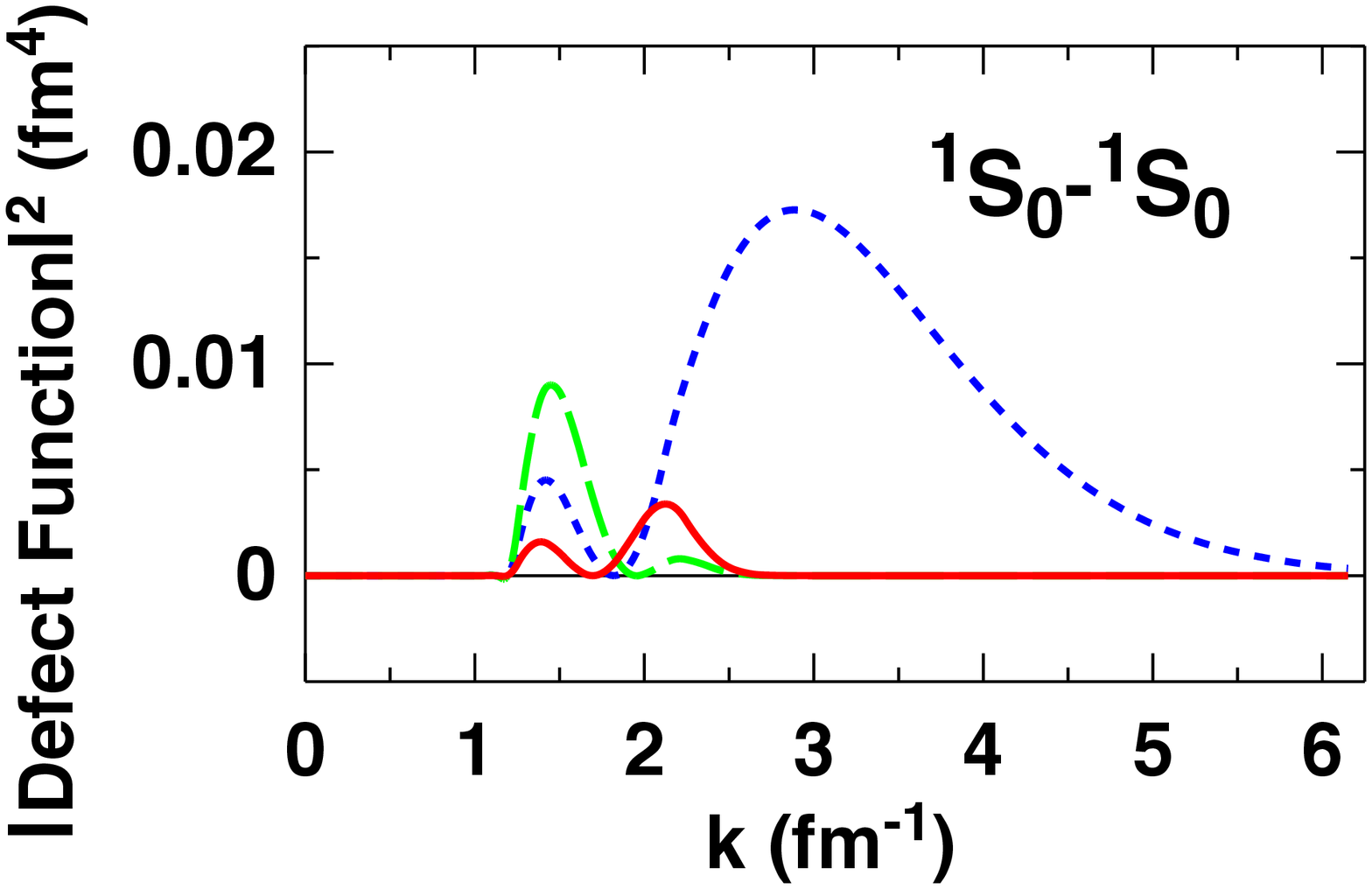}}
\vspace*{-3.0cm}
\caption{(Color online)                                              
Blue (dotted): Magnitude squared of the defect function 
from the DBHF calculations together with the Bonn B potential; green (dashed): prediction 
with the chiral two-body interaction only; red (solid): prediction with two- and 
three-body chiral interactions. Symmetric nuclear matter with Fermi momentum equal to 1.4 fm$^{-1}$. 
}
\label{2b3b}
\end{figure}

In Fig.~\ref{2b3b}, the blue (dotted) curve shows the predictions 
from Bonn B we already discussed; the green (dashed) curve displays the prediction 
with the chiral two-body interaction only; finally, the red (solid) curve is obtained with two- and 
three-body chiral interactions. In all cases, the Fermi momentum is equal to 1.4 fm$^{-1}$. 

Short-range correlations with chiral or meson-theoretic interactions can be dramatically
different. In particular, the defect functions obtained with chiral potentials is restricted to smaller momenta than those derived from
meson theory. This is due to the different cutoffs discussed above.

Clearly, the chiral 3NF contributes to the tensor force. For instance, 
for the $^3S_1-^3D_1$ transition near normal density, it increases the probability amplitude 
around 1.5-2 fm$^{-1}$ by about 30\%. 

In Table~\ref{tab2} we compare 
the wound integral in symmetric matter for the three approaches considered in Fig.~\ref{2b3b}. 

\begin{table}[!htb]                
\centering \caption                                                    
{The wound integral $\kappa$ in SNM for the three calculations shown in Fig.~\ref{2b3b}. 
The total density is equal to 0.185 fm$^{-3}$.
} 
\vspace{5mm}
\begin{tabular}{|c|c|}
\hline
Theoretical approach &   $\kappa$    \\ 
\hline
  Bonn B + DBHF &  0.130             \\                                       
\hline     
Chiral $NN$ (2NF) & 0.075          \\
\hline
Chiral $NN$ + 3NF & 0.099          \\
\hline
\hline
\end{tabular}
\label{tab2} 
\end{table}

We conclude this section with some comments on the experimental side of the SRC discussion presently
going on in the literature. 
Inclusive electron scattering measurements at high momentum transfer, on both light and heavy nuclei,
have been analyzed with the purpose of extracting information on SRC.\cite{CLAS,src,Pia+} 
In a suitable range of $Q^2$ and $x_B$, the cross section is factorized in order to single out the probability
of a nucleon to be involved in SRC, either two-body or three-body. When extended to nuclear matter, 
this probability should be comparable to the wound integral we discussed above.    

Nuclear scaling and the plateaus seen in inclusive scattering cross section ratios\cite{CLAS} 
are due to the dominance of SRC for momenta above approximately 2 fm$^{-1}$. In the same region,
the momentum distribution in a nucleus relative to the one in the deuteron becomes almost flat, so that those
distributions simply scale with $A$.              

The probabilities mentioned above are a manifestation of the off-shell nature 
of the potential, which cannot be determined uniquely from $NN$ elastic data and is not an observable. As is well 
known, interactions may differ dramatically in their off-shell behaviour while remaining phase-equivalent.                                           
In a recent paper,\cite{src15} we explored to which extent modern, {\it non-phenomenological} interactions are 
consistent with the information as extracted from $A(e,e')X$ measurements. We took the deuteron as                                      
the simplest system where off-shell behaviour can be explored. We recall that the high-momentum
part of the momentum distribution shows similar features in nuclei with $A$=2 to 40.\cite{Alv13} Thus, 
the deuteron offers representative features. Furthermore, deuteron SRC probabilities are a crucial element 
in the estimation of SRC probabilities in heavier nuclei as obtained in Ref.\cite{CLAS}. 
                  
Characteristic differences exist between meson-theoretic potentials using 
fully relativistic one-pion exchange amplitudes (that is, non-local tensor forces) and those which use 
static one-pion-exchange. In Ref.\cite{src15} we found that the SRC probability in the deuteron predicted with 
a high-quality, non-local meson-exchange potential is roughly 25\% below the value                  
cited in Ref.\cite{CLAS} and used to evaluate absolute probabilities in heavier nuclei.                           
A qualitatively similar disagreement exists for the wound integral in nuclear matter. More precisely, 
conventional non-local potentials are known to predict about 10-15\% for the wound integral 
(at normal density),\cite{FS14} whereas a value of about 25\% is cited from extrapolation from heavy nuclei 
to nuclear matter.\cite{Pia13} This discrepancy should be kept in mind when interpreting 
the empirical information.

\section{Microscopic Optical Model\label{sect4}}
The optical model is a crucial component in nuclear reaction
studies, mainly because it determines the cross section for nuclear
scattering and the formation of compound nuclei in the initial stage of a
reaction. Furthermore, it supplies the transmission coefficients for branching
into the various final states.\cite{M.T.Pigni2011} Many observables
such as the elastic scattering angular distribution, analyzing
power, spin rotation function, and more can be derived through the
optical model.

Many attempts have been made to derive a Microscopic Optical Potential (MOP) within the framework of a microscopic many-body theory from a realistic $NN$
interaction. Pioneering work along this line has been presented by Mahaux and coworkers\cite{Jeu77}
who evaluated the nucleon self-energy in nuclear matter as a function of density and energy
in a BHF approximation and identified the resulting complex single-particle
potential with the MOP for finite nuclei using LDA. The so-called $G$-folding method developed by Amos et al.\cite{K.Amos2000} is also based
on a realistic $NN$ interaction and uses a local density approximation to account
for the medium dependence of the effective interaction. In this case, however, it is the $NN$
interaction which is evaluated by solving the Bethe-Goldstone equation in nuclear matter and
then employed in a folding calculation to evaluate the MOP for finite nuclei. The $G$-folding
approach has been applied very successfully to reproduce differential cross sections and spin
observables for many nuclei from $^{6}$Li to $^{238}$U without adjustable parameters.\cite{K.Amos2000,P.J.Dortmans1998,deb01,brown00,P.K.Deb2005,A.Lagoyannis2001,M.Dupuis2006,M.Dupuis2008} 

The Mahaux scheme as well as the $G$-folding method are based on a non-relativistic approach, which fails to reproduce the saturation properties of nuclear systems.
The energy-dependence of the MOP originates from the energy-dependence of the effective interaction $G$ calculated
for nuclear matter in a non-relativistic Brueckner Hartree Fock approximation. 

To derive a MOP based on the relativistic DBHF approach\cite{RRXU2012,RRXU2016} the relativistic structure of the nucleon self-energy has been considered
in asymmetric nuclear matter. Generalizing Eq.~(\ref{subsec:SM;eq:self1}) we obtain
\begin{eqnarray}
     \Sigma^\tau(k,E,k_F,\alpha) = \Sigma^\tau _s(k,E,k_F,\alpha) - \gamma_0\Sigma_0^\tau (k,E,k_F,\alpha) \;  \label{eq5}\\
     \nonumber + \bm{\gamma}\cdot
     \textbf{k}\Sigma_v^\tau (k,E,k_F,\alpha)~,
\end{eqnarray}
where the label $\tau$ refers to the self-energy of protons or neutrons and 
$$
\alpha = \frac{\rho_n - \rho_p}{\rho}
$$
defines the asymmetry parameter with $\rho_n$, $\rho_p$ and $\rho$ indicating the neutron density, proton density and total density, respectively. The energy variable $E$ is normalized in such a way that $E=0$ corresponds to the Fermi energy at density $\rho$ and asymmetry $\alpha$ under consideration. This implies that one obtains imaginary components for $E>0$. The analysis 
of Ruirui Xu {\it et al.}\cite{RRXU2016} is based on the Bonn B potential\cite{machl89} and employs the subtracted T-matrix  representation
as described in Ref.\cite{Dalen10} to extract the Dirac components of the self-energy.

Using the definitions
\begin{eqnarray}
      U_s^\tau=\frac{\Sigma_s^\tau-\Sigma_v^\tau}{1+\Sigma_v^\tau},~~U_0^\tau=\frac{-\Sigma_0^\tau+\varepsilon\Sigma_v^\tau}{1+\Sigma_v^\tau}\,,~~ \label{eq11}
\end{eqnarray}
the Dirac equation for the scattering of a particle with incident kinetic energy $E$ and corresponding relativistic energy $\varepsilon$ = $\textit{E}+\textit{M}$
can be written as 
\begin{eqnarray}
      \left[\vec{\alpha}\cdot\vec{p}+\gamma_0(\textit{M}+U_s^\tau)+U_0^\tau \right]
      \Psi^\tau=\varepsilon\Psi^\tau~ . \label{eq10}
\end{eqnarray}
\begin{figure*}[htbp]
\centerline{\includegraphics[width = 9.0cm]{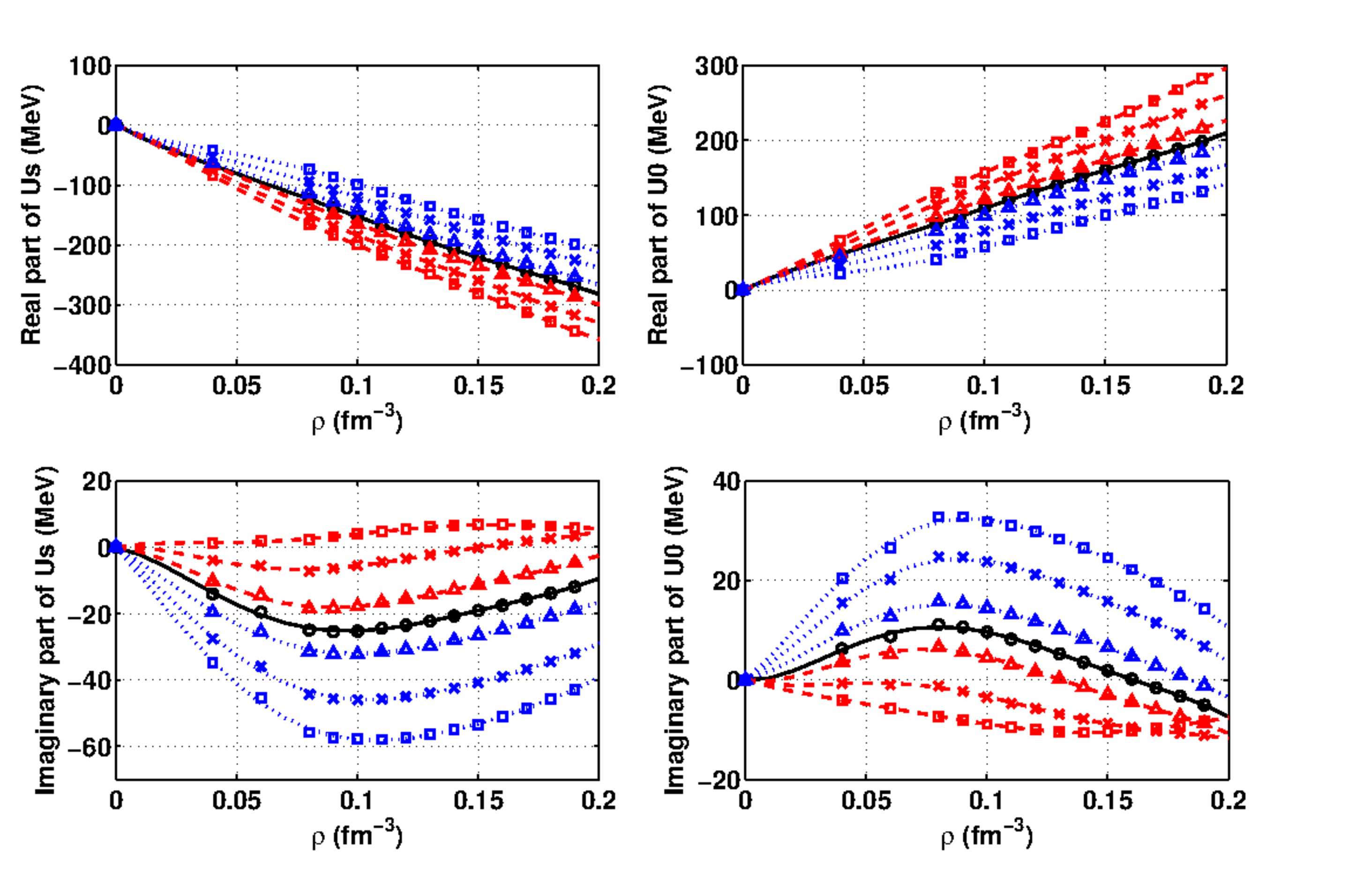}}
\caption{ (color online) Example for the real and imaginary part of the scalar ($U_s$) and
vector ($U_0$) components of the Dirac potential as a function of density for nucleons with
an incident particle energy of 90 MeV. The circles,
triangles, ``x" marks, and squares represent the calculated (adjusted) values for isospin asymmetries
$\alpha =$ 0.0, 0.2, 0.6 and 1.0, respectively. The connecting solid line shows the polynomial
 interpolation in
the case of symmetric matter, while the dashed and dotted lines visualize the corresponding
interpolations for the neutron- and proton-potentials, respectively.}
\label{fig2mop}
\end{figure*}

Results for the real and imaginary parts of the Dirac components $U_s^\tau$ and $U_0^\tau$ are presented in Fig.~\ref{fig2mop}. 
Note that DBHF calculations for homogeneous infinite matter yield reliable results only
for densities $\rho >$ 0.08 fm$^{-3}$. The procedure to derive
self-consistent DBHF results does typically not converge at lower
densities. This reflects the situation that homogeneous nuclear
matter is unstable at such low densities with respect to the
formation of an inhomogeneous density profile containing nuclear
clusters. 
To derive the optical model potential for finite nuclei,
however, we also need predictions at densities $\rho<$ 0.08 fm$^{-3}$.
Therefore we have to extrapolate the results to these low densities
with the natural constraint that the Dirac potentials $U_s^m$ and
$U_0^m$ vanish at $\rho = 0.$

Note that the analysis of the relativistic structure of  the self-energy in neutron-rich asymmetric nuclear matter yields a more attractive
real part of the scalar potential for the protons than for the neutrons, which leads to a smaller Dirac mass for the protons. The real part
of the vector potential, on the other hand, is more repulsive for the protons. The opposite behaviour is observed for the imaginary part. 
Here larger values are typically obtained for the neutrons.

In order to calculate observables for nucleon-nucleus scattering, 
the Dirac equation, Eq.~(\ref{eq10}), is typically reduced to a Schr\"{o}dinger type
equation  by eliminating the lower components of the Dirac spinor with a 
 standard procedure. The equation for the upper components of the wave
function is transformed into:
\begin{eqnarray}
\left[-\frac{\nabla^2}{2{\varepsilon}}+V^\tau_{cent}+V^\tau_{s.o.}(r)\vec{\sigma}\cdot\vec{\textbf{\L}}+V^\tau_{Darwin}(r)
\right]\varphi(\textbf{r}) \\
      \nonumber =\frac{\varepsilon^2-\textit{M}^2}{2\varepsilon}\varphi(\textbf{r}), \label{eq13}
\end{eqnarray}
where $V^\tau_{cent}$, $V^\tau_{s.o.}$ and $V^\tau_{Darwin}$ represent
the Schr\"{o}dinger equivalent central, spin-orbit
and Darwin potentials, respectively.
The potentials in Eq.~\ref{eq13} are obtained from the scalar $U_s$ and vector $U_0$ potentials as
\begin{eqnarray}
      \nonumber V^\tau_{cent}=\frac{M}{\varepsilon}U_s^\tau+U_0^\tau+\frac{1}{2\varepsilon}[U_s^{\tau 2}-(U_0^\tau+V_c)^2], \label{eq14}
\end{eqnarray}
\begin{eqnarray}
      V_{s.o.}^\tau=-\frac{1}{2{\varepsilon}rD^\tau(r)}\frac{dD^\tau(r)}{dr}, \label{eq15}
\end{eqnarray}
\begin{eqnarray}
     \nonumber V_{Darwin}^\tau=\frac{3}{8{\varepsilon}D^\tau(r)}\left[\frac{dD^\tau(r)}{dr}\right]^2-\frac{1}{2{\varepsilon}rD^\tau(r)}\frac{dD^\tau}{dr} \\
     \nonumber -\frac{1}{4{\varepsilon}D^\tau(r)}\frac{d^2D^\tau(r)}{d^2r},\label{eq16}
\end{eqnarray}
where $V_c$ is the Coulomb potential for a charged particle and $D$
denotes a  quantity defined as
\begin{eqnarray}
       D^\tau(r)=M+\varepsilon+U_s^\tau(r)-U_0^\tau(r)-V_c. \label{eq17}
\end{eqnarray}
The central part of the Schr\"odinger equivalent potential $V_{cent}$ is dominated by the sum of the strongly attractive scalar part of the Dirac potential $U_s$ and the strongly repulsive vector component $U_0$ which cancel each other to a large extent as discussed previously. Typical is also the energy dependence contained in the correction terms of second order in $U_s$  and $U_0$. The Schr\"odinger spin-orbit term $V_{s.o.}$, on the other hand, contains the difference of $U_s$ and $U_0$ in the denominator. This difference is large and leads to a larger spin-orbit splitting than usually predicted in non-relativistic approaches.\cite{dalenrev} 

The radial dependence of the Dirac potentials $U_s$ and $U_0$ can be determined from the corresponding quantities in nuclear matter using LDA, which is based on proton $\rho_p(r)$ and neutron density distributions $\rho_n(r)$ or the corresponding total density $\rho(r)$ and asymmetry $\alpha(r)$ for the target nucleus under consideration: 
\begin{eqnarray}
      U^\tau_{LDA}(r,E)=U^\tau_{NM}(k,E,\rho(r),\alpha(r))\,. \label{eq9}
\end{eqnarray}
The LDA in this simple form makes sense if one can ignore the range of the $NN$ interaction. Finite range effects are approximated by the Improved Local Density Approximation (ILDA),\cite{Jeu77}  
\begin{equation}
       U^\tau_{ILDA}(r,E)=(t^\tau\sqrt{\pi})^{-3}
      \int{U^\tau_{LDA}(r',E)exp(-|\vec{r}-\vec{r'}|^2/t^{\tau}2)d^{3}r'}~, \label{eq18}
\end{equation}
where $t^\tau$ are effective range parameters for proton and neutron scattering.

If the Dirac potentials $U_s$ and $U_0$ are determined from DBHF calculations of asymmetric nuclear matter and the nucleon density distributions are fixed e.g. by using the results from Hartree-Fock-Bogoliubov calculations with Gogny D1S force,\cite{Hilaire2007} there remain only the two finite range parameters, $t^p$ and $t^n$, to adjust the MOP. These two parameters and the extrapolations of Dirac potentials at low densities (see Fig.\ref{fig2mop}) have been fitted to reproduce the experimental data of proton and neutron scattering from $^{48}$Ca an $^{208}$Pb, for which a large number of experimental data  are available\cite{RRXU2016}. They are both double-magic nuclei and represent proper examples to cover a good range of isospin assymetry.  Xu {\it et al.} \cite{RRXU2016} demonstrate, that using $t_n$ = 1.35 fm and $t_p$ = 1.45 fm the resulting potential CTOM (China Nuclear Data Center and T\"ubingen University Optical Model) not only yields good results for the data to fit but also for other nuclear targets.

\begin{figure}[htbp]
\centerline{\includegraphics[width = 9.0cm]{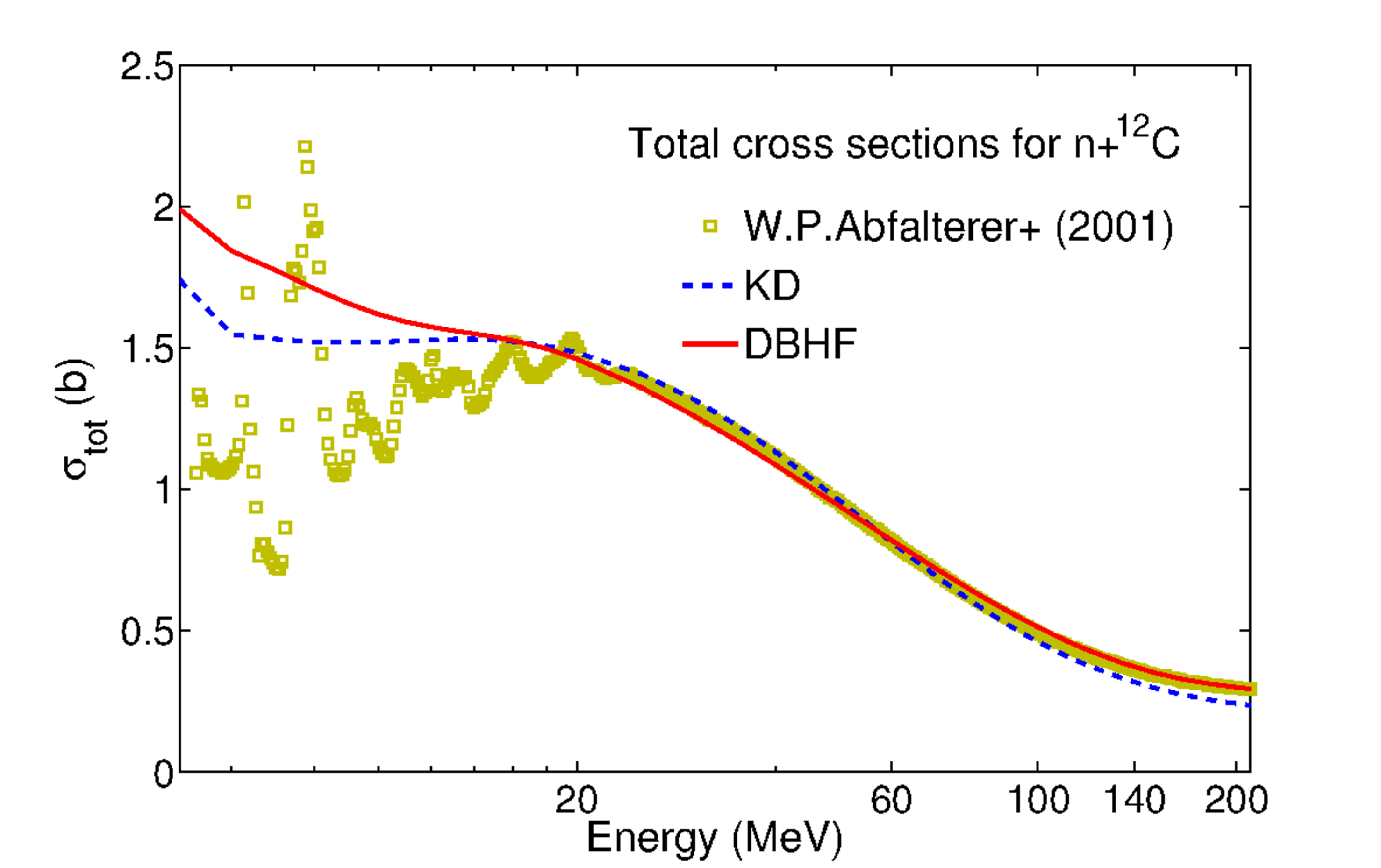}}
\caption{(color online) Comparison of predicted neutron total cross section (solid line) and experimental data (point) and KD calculation (dashed line) for $n$ +$^{12}$C.}
\label{fig3}
\end{figure}

\begin{figure*}[htbp]
\centerline{\includegraphics[width = 9cm]{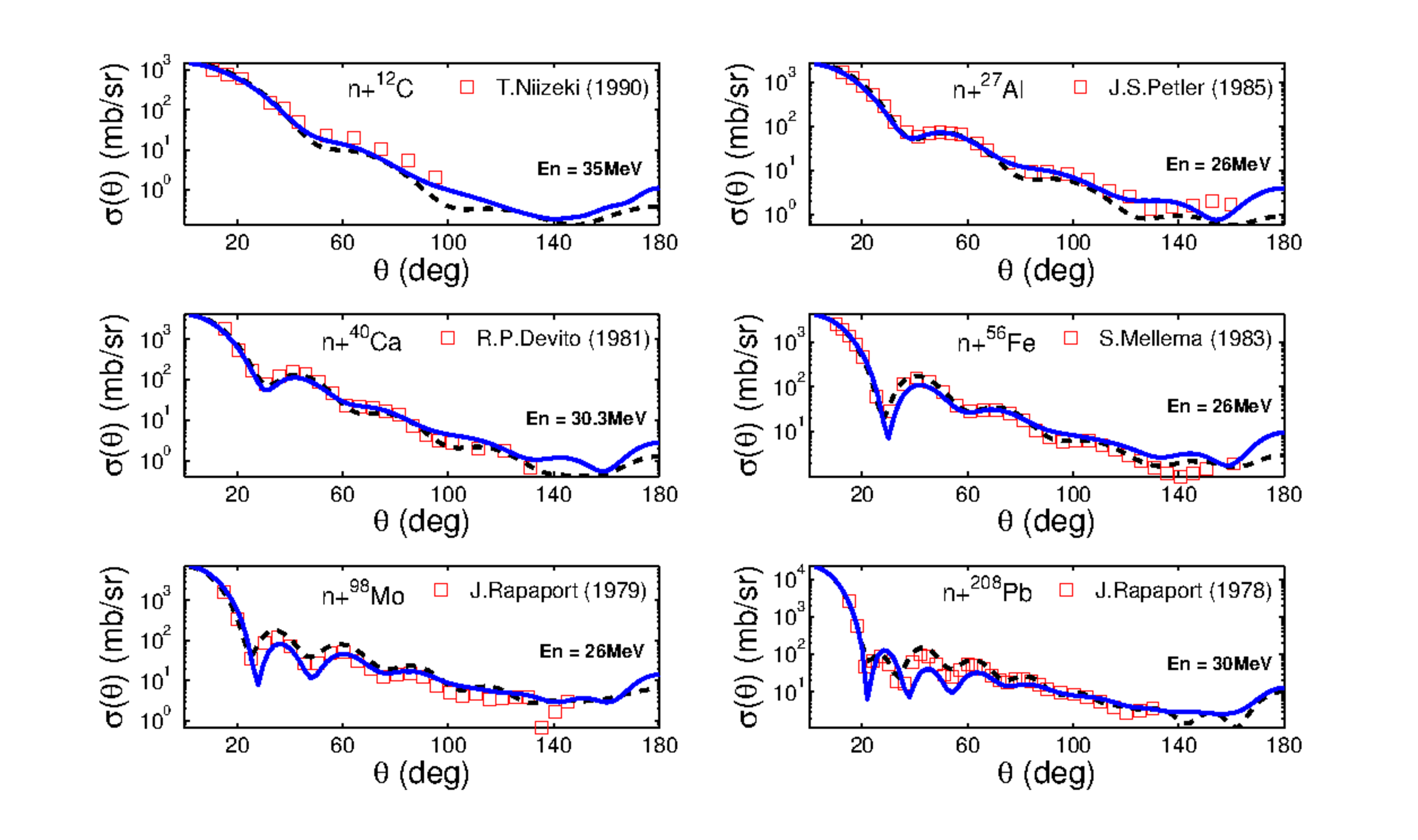}}
\caption{(color online) Comparisons of angular distributions for $n$
+ $^{12}$C,$^{27}$Al,$^{40}$Ca,$^{56}$Fe,$^{98}$Mo and $^{208}$Pb at
incident neutron energy around 30MeV. The dashed line indicates the
results from KD potential and the solid line denotes the DBHF
prediction.} \label{fig6}
\end{figure*}

 As typical results, we present in Fig.~\ref{fig3} the total cross section for neutron scattering on $^{12}$C as a function of projectile energy, whereas the angular distribution for
neutron scattering on various targets at an incident energy of 30 MeV is shown in Fig.~\ref{fig6}. The results of the MOP derived from the DBHF calculations are represented by solid lines and compared to the experimental data and the predictions of the phenomenological Koning-Delaroche (KD) global potential.\cite{Kon03} 

It is obvious that a Optical Model Potential can only describe global features of nucleon scattering. The scattering at small energies is dominated by the surface excitation modes, which are specific to each nucleus. Therefore, both the KD optical model as well as the MOP based on the DBHF approach reproduce only the general trend of the total cross section at energies below 20 MeV (see Fig.\ref{fig3}). For larger energies, however, both approaches reproduce the total cross section as well as the angular distributions very nicely.

\begin{figure*}[htbp]
\centerline{\includegraphics[width = 9.0cm]{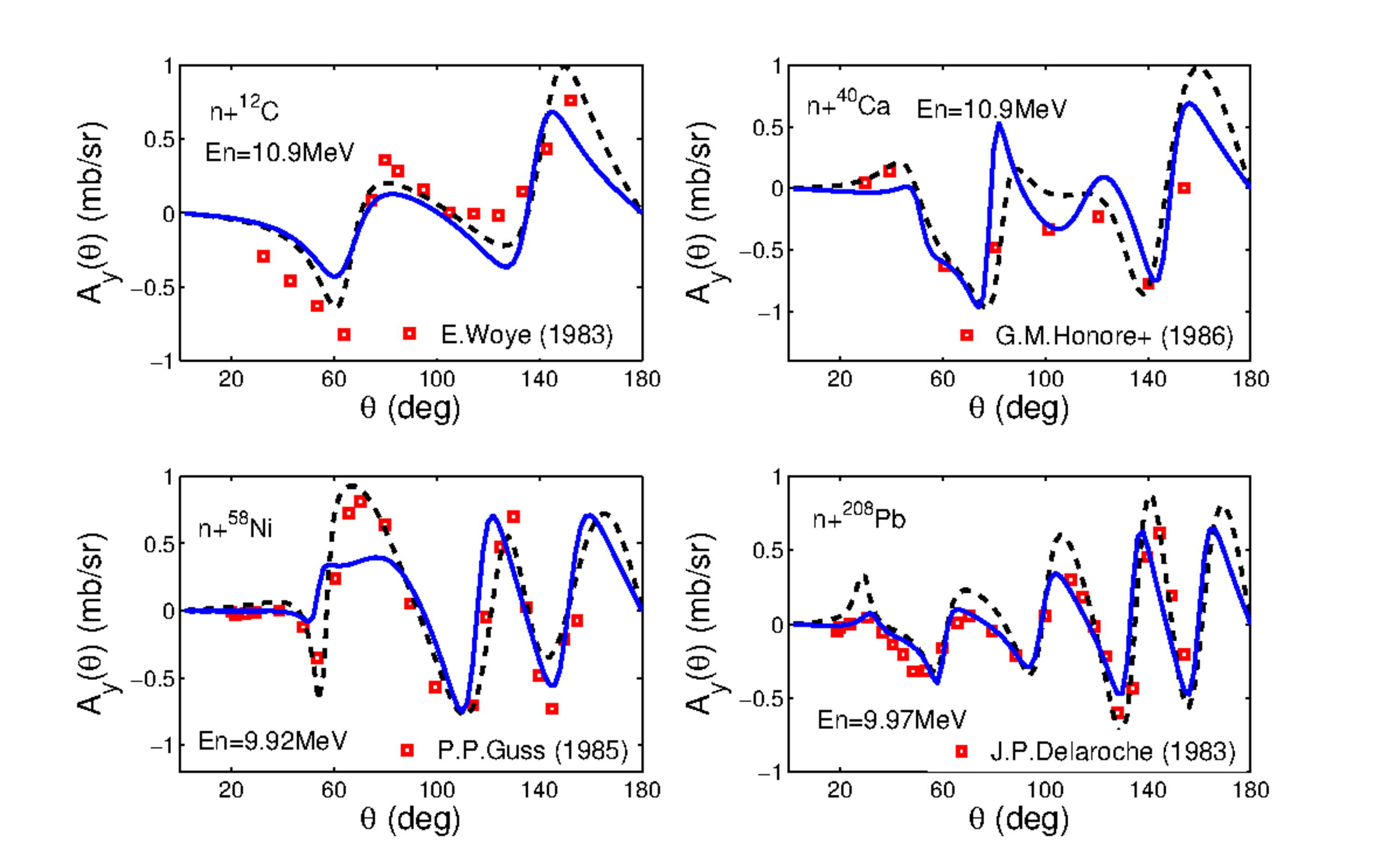}} 
\caption{
Comparisons of analyzing power for $n$ + $^{12}$C, $^{40}$Ca,
$^{58}$Ni, and $^{208}$Pb at incident neutron energy around 10MeV.
The dashed line indicates the results from KD potential and the
solid line denotes the MOP derived from DBHF.}
\label{fig14}
\end{figure*}

The same is also true for the analyzing power $A_y$ as a function of scattering angle $\theta$ (see Fig.~\ref{fig14}). The analyzing power provides a very sensitive test of the spin-orbit term. This spin-orbit term emerges directly from the DBHF analysis (see Eq.~\ref{eq15}), whereas a separate fit procedure is required in a phenomenological potential.

This analysis demonstrates that the DBHF approach is not only able to describe the saturation properties of nuclear matter, but also provides detailed information on nucleon-nucleus scattering observables without the adjustment of many parameters, which are required in a phenomenological approach to determine a global optical model potential.  

\section{Relativistic effects and three-nucleon forces\label{sect5}}

It has already been discussed in Section~\ref{sect2} that the main feature of the DBHF approach not included in a non-relativistic BHF calculation, namely the enhancement of the small component of the Dirac spinor in the nuclear medium, can be expressed in terms of the $Z$-diagram displayed in Fig.~\ref{3b} and the corresponding 3NF          
(see Eq.~\ref{eq:3-v}):
\begin{equation}
V_{ijk} =  U\,V_{ijk}^R\,.\label{eq:3-vp}
\end{equation}
Such a 3NF can be implemented in the many-body calculation in terms of a density-dependent two-nucleon interaction, $V^{eff}$, if the single-particle energies are defined using the
Landau prescription
\begin{equation}
\varepsilon_i = \frac{\partial}{\partial \rho_i} E(\rho)\,,\label{eq:landaudp}
\end{equation}
implying that rearrangement terms due to the density dependence of $V^{eff}$ are taken into account. Note, however,  that when using the Brueckner $G$-matrix for the two-nucleon interaction in the medium, the density dependence of $V^{eff}$ is due to the implementation of the 3NF as well as the presence of the Pauli operator in the Bethe-Goldstone equation, Eq.~(\ref{eq:bg}), and the dependence of $G$ on the starting energy.

In fact, applying the Landau definition of the quasiparticle energies of Eq.~(\ref{eq:landaudp}) to the BHF energy functional  
$$
E = \sum_{i<F} t_i + \frac12\sum_{i,j<F} \langle ij\vert G(\omega = \varepsilon_i + \varepsilon_j) \vert ij\rangle\rho_i\rho_j \; , 
$$
one obtains the BHF terms of Eq.~(\ref{eq:bhf1}) plus two additional terms, the starting energy rearrangement term $\Delta U_i^\omega$ and the Pauli rearrangement term $\Delta U_i^Q$, which are due to the dependence of $G$ on starting energy $\omega$ and Pauli operator $Q$.

The starting energy rearrangement term can be written as: 
\begin{eqnarray}
\Delta U_i^\omega & = & \sum_{j,k}\rho_j\rho_k \langle j,k\vert \frac{\partial G}{\partial \omega}\vert j,k\rangle \frac{\partial\varepsilon_j}{\partial \rho_i} \nonumber\\
& = & \sum_{j,k}\rho_j\rho_k \langle j,k\vert \frac{\partial G}{\partial \omega}\vert j,k\rangle \langle i,j\vert G\vert i,j\rangle\,. \label{eq:startre1}
\end{eqnarray}
The second line of this equation is obtained by substituting $\varepsilon_j$ in the first line by the corresponding BHF definition of the single-particle energy. Note that adding $\Delta U_i^\omega$ to the BHF definition of the single-particle energy leads to
\begin{eqnarray}
\varepsilon_i^{RBHF} & = & \varepsilon_i + \Delta U_i^\omega \label{eq:rbhf1}\\
& = &  \langle i \vert \hat t \vert i \rangle + \sum_{j} \langle ij \vert G(\omega=\varepsilon_i + \varepsilon_j)\vert ij \rangle P_j\,,\nonumber\\
\end{eqnarray}
which implies that we have replaced the single-particle density $\rho_j$ in Eq.~(\ref{eq:bhf1}) by
\begin{equation}
P_j = \rho_j \left[ 1 + \sum_k \rho_k \langle j,k\vert \frac{\partial G}{\partial \omega}\vert j,k\rangle \right]\,.
\label{eq:occupation}
\end{equation}
This expression for $P_j$ typically yields values of the order of $0.8$ to $0.9$ and is often interpreted as a partial occupation of states $j$ below the Fermi energy. The approximation Eq.~(\ref{eq:rbhf1}) represents the leading terms of the so-called Renormalized BHF approach (RBHF).\cite{negele,eigenearb} We will use this acronym in the following.

The Pauli rearrangement term can be written as
\begin{equation}
\Delta U_i^Q =- \sum_{j,k,l}\rho_j\rho_k \left\vert\langle j,k\vert G \vert i,l\rangle \right\vert^2\frac{1-\rho_l}{\varepsilon_j+\varepsilon_k-\varepsilon_i-\varepsilon_l}\,,\label{eq:paulir}
\end{equation}
and corresponds to the term of second order in $G$ in the hole-line expansion of the self-energy. Calculations including Pauli and starting energy rearrangement terms are denoted as density-dependent Hartree-Fock (DHF) calculations and employ single-particle energies of the form
\begin{equation}
\varepsilon_i^{DHF}  =  \varepsilon_i + \Delta U_i^\omega + \Delta U_i^Q\,.\label{eq:dhf1}
\end{equation}

All the calculations presented here have been performed using the proton-neutron part of the charge-dependent Bonn interaction (CD Bonn).\cite{cdbonn}  

\begin{figure*}[htb]
\begin{center}
\includegraphics[width=0.8\textwidth]{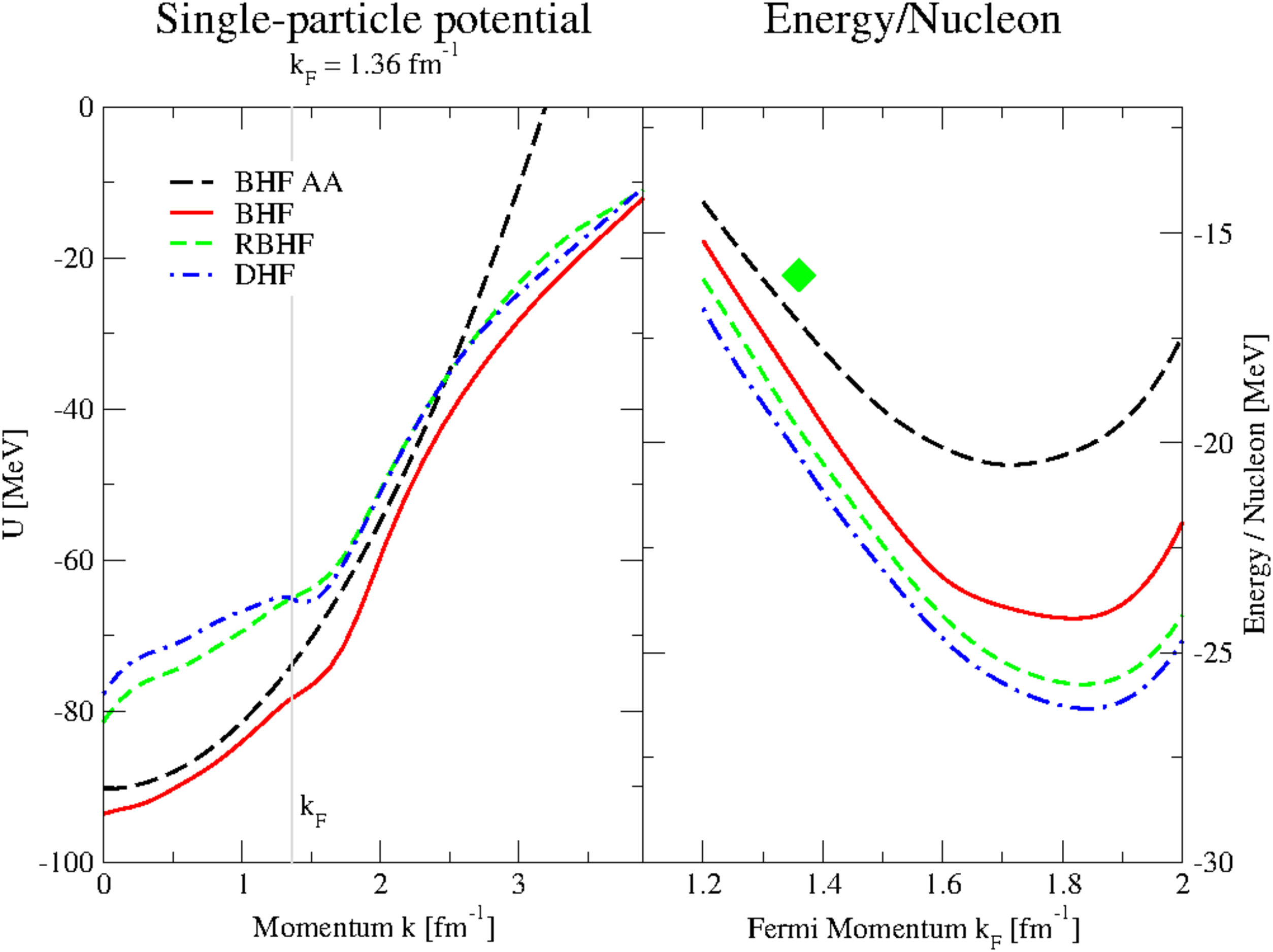}
\caption{ Results for symmetric nuclear matter calculations using the pn interaction of the CD Bonn potential. The left panel presents results for the single-particle potential $U(k)$ assuming a Fermi momentum $k_F$ of 1.36 fm$^{-1}$, which corresponds to the empirical saturation densities.  Results are displayed for the BHF approximation using the angle-average in the Bethe-Goldstone equation (BHF AA) and BHF, RBHF (see eq.(\protect{\ref{eq:rbhf1}})) and DHF (see eq.(\protect{\ref{eq:dhf1}})) calculations solving the Bethe-Goldstone equation without this approximation. The right panel shows the corresponding results for the energy per nucleon calculated at various Fermi momenta. }
\label{fig:nucmat1}
\end{center}
\end{figure*}

Results of conventional BHF calculations for symmetric nuclear matter, using the angle-average approximation for the Pauli operator in the Bethe-Goldstone equation and a conventional quadratic parameterization of the single-particle potential are presented by the dashed line, labeled BHF AA, in Fig.~\ref{fig:nucmat1}. The single-particle potential, which is displayed in the left panel of this figure for a Fermi momentum $k_F$ of 1.36 fm$^{-1}$, (corresponding to the empirical saturation density), reflects the quadratic parameterization adjusted to reproduce the BHF single-particle potential $U(k)$ for momenta $k$ below the Fermi momentum and extended to momenta above $k_F$.  

The calculated binding energy per nucleon of the BHF AA calculations, shown in the right panel of Fig.~\ref{fig:nucmat1}, yield a minimum at about twice the empirical saturation density and an energy around -20 MeV, which indicates a much stronger attraction than required by the empirical value of -16 MeV.

The consistent treatment of the two-particle propagator in the Bethe-Goldstone equation, Eq.~(\ref{eq:bg}), avoiding the angle-average of the Pauli operator and using a consistent single-particle spectrum, leads to quite different results, as can be observed from a comparison of the BHF predictions, see  solid red curves in Fig.~\ref{fig:nucmat1}, with those from the BHF AA predictions. As discussed by Schiller et al.,\cite{schillerq} these differences can mainly be attributed to the definition of the single-particle potential. As can be seen from the left panel of Fig.~\ref{fig:nucmat1}, the single-particle energies used to define the propagator of the Bethe-Goldstone equation are quite similar for momenta below $k_F$. For particle states with momenta above $k_F$, however, the calculated BHF energies are more attractive than described by the quadratic parameterization of the BHF AA approach.

The corresponding differences in the two-particle propagator then lead to more attractive $G$-matrix elements in the BHF  approach, which, in turn, yield more binding energy. This can be seen from the energy {\it vs.} density curves, presented in the right panel of Fig.~\ref{fig:nucmat1}. The BHF calculations yield a saturation point with even larger binding energy (-24 MeV) than the BHF AA approach at a larger saturation density.

\begin{figure*}[htb]
\begin{center}
\includegraphics[width=0.8\textwidth]{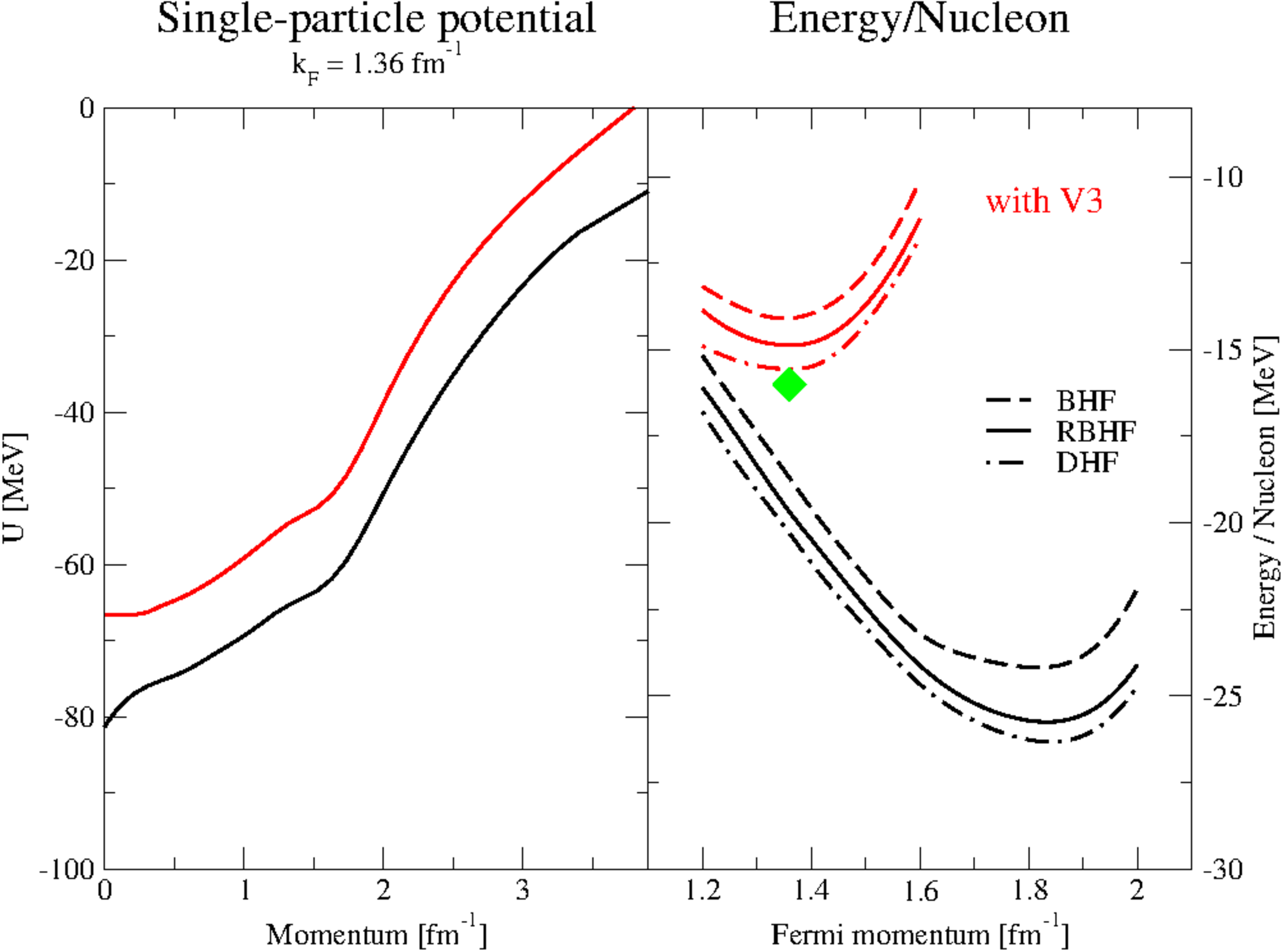}
\caption{(Color online)  Results for symmetric nuclear matter calculations using  the CD Bonn potential with (red curves) and without (black curves) inclusion of the $3N$ potential. The left panel presents results for the single-particle potential $U(k)$ with the RBHF approximation assuming a Fermi momentum $k_F$ equal to 1.36 fm$^{-1}$. The right panel shows results for the energy per nucleon calculated at various Fermi momenta with the BHF, RBHF, and DHF approximations. }
\label{fig:nucmat2}
\end{center}
\end{figure*}

All calculations discussed so far have been performed with just one realistic $NN$ interaction, namely the CD Bonn potential. We find that the predicted saturation points are part of the Coester band discussed in the Introduction. This is true for the BHF approach, where the rearrangement terms provide only a shift along the Coester band. As one of the main goals of this investigation is to explore the possibility to simulate relativistic effects in terms of a $3N$ potential, we considered the $3N$ interaction of  Eq.~(\ref{eq:3-vp}) and adjusted the parameter $U$ to reproduce the empirical saturation point.

Results of such calculations are displayed in Fig.~\ref{fig:nucmat2}. It is worth noting that one can obtain a good description of the empirical saturation point by adjusting only one parameter, whereas two or more parameters in the $3N$ interaction have typically been found necessary for a fit of corresponding quality.

\begin{table*}[tb]
\begin{tabular}{c|ccc|ccc|c}
$^{16}$O& \multicolumn{3}{c|}{$NN$ only} & \multicolumn{3}{c|}{with $3N$}&   Exp.\\
& \multicolumn{1}{c}{BHF} & \multicolumn{2}{c|}{RBHF}&  \multicolumn{1}{c}{BHF} & \multicolumn{2}{c|}{RBHF}& \\
\hline
& $\varepsilon$ [MeV] & $\varepsilon$ [MeV] & $P$  & $\varepsilon$  [MeV] & $\varepsilon$  [MeV]& $P$ & $\varepsilon$\\
&{Protons}&&&&& \\
$s_{1/2}$& -58.19 &-48.69&0.892 &-44.76  &-36.88  &0.917 &-44 $\pm$ 7\\
$p_{3/2}$& -27.05 &-20.93&0.897 &-20.22 &-14.82 &0.840 & -18.45\\
$p_{1/2}$& -20.02 &-16.25&0.871 &-16.50 &-12.27 &0.824 & -12.12\\
&{Neutrons}&&&&&\\
$s_{1/2}$& -62.22 &-52.07&0.892 & -48.36&-39.67 & 0.918& -47\\
$p_{3/2}$& -30.98 &-24.19&0.901 & -23.71&-17.60 &0.846  & -21.84\\
$p_{1/2}$& -23.83 &-19.44&0.875 & -20.00&-15.05 & 0.829 & -15.66\\
\hline
E/A [MeV]
& -6.08 &  -6.57 & &-4.61 & -5.22&  &-7.98 \\ 
$R_c$ [fm] & 2.35 & 2.45 && 2.59 &2.66 & &2.74 \\
\end{tabular}
\caption{Results for $^{16}$O using  BHF and RBHF approximation without ($NN$ only) and with inclusion of the $3N$ interaction (with $3N$). Values of single-particle energies ($\varepsilon$) occupation probabilities ($P$, see Eq.~(\protect{\ref{eq:occupation}})) are listed for the occupied states as well as the energy per nucleon (E/A) and the radius of the charge distribution  ($R_c$). \label{VergleichO1}}
\end{table*}

The focus of this investigation is to explore whether a BHF calculation with a parameterization of the Dirac effects in terms of a $3N$ interaction can provide a good description of the nuclear matter saturation point as well as the bulk properties of finite nuclei. For that purpose we performed BHF and RBHF calculations for the closed shell nuclei $^{16}$O and $^{40}$Ca using the same $NN$ and $3N$ interactions as just described for nuclear matter.

\begin{figure*}[htb]
\begin{center}
\includegraphics[width=0.8\textwidth]{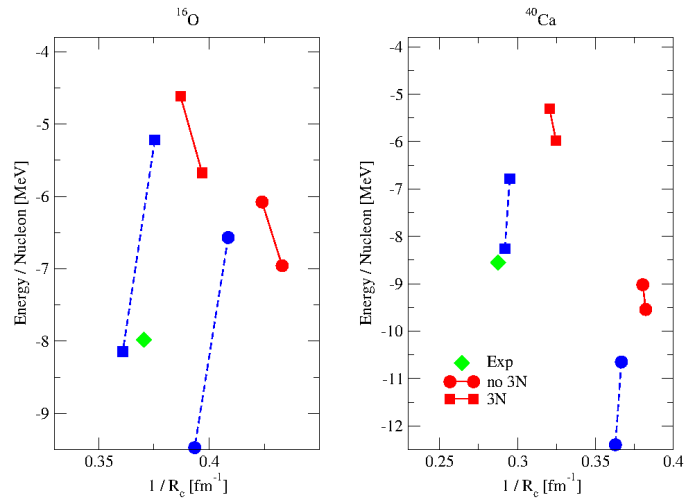}
\caption{(Color online)  Results for the energy per nucleon and the charge radius ($R_c$) for $^{16}$O (left panel) and $^{40}$Ca are presented in a plot of energy {\it vs.} $1/R_c$ to enable the comparison with the corresponding figures for nuclear matter (Fig.~\protect{\ref{fig:nucmat1}} and Fig.~\protect{\ref{fig:nucmat2}}).  Results referring to BHF calculations with different choices for the particle spectrum in the Bethe-Goldstone equation are connected by a red solid line, those of RBHF calculations are connected by a blue dashed line. Results of calculations with $NN$ interaction only, with inclusion of the $3N$ term, and with inclusion of the $3N$ term {\it via} a global density approximation are indicated with a dot, a box, and a cross, respectively. The experimental result is shown by the green diamond. }
\label{fig:oca1}
\end{center}
\end{figure*}

Results of BHF calculations for  $^{16}$O using only the CD-Bonn potential are presented in Table~\ref{VergleichO1}. One finds that the predicted energy per nucleon (-6.08 MeV) is considerably less than the experimental value (-7.98 MeV) and the calculated charge radius, $R_c$, is much smaller (2.35 fm) than the empirical value of 2.74 fm. In Fig.~\ref{fig:oca1} we wish to visualize this result as the counterpart, for a nucleus, of the nuclear matter saturation point as shown, for instance, in the right panel of Fig.~\ref{fig:nucmat2}. In a plot of the energy {\it vs.} the inverse of the charge radius, we indicated this result by a red dot, see Fig.~\ref{fig:oca1}. In fact, it is the upper of the two red dots, connected by a solid line in the left panel  of this figure.

When comparing with the empirical data, represented by the green diamond, we see that the situation for $^{16}$O is quite different than for nuclear matter (see Fig.~\ref{fig:nucmat2}). In both cases the BHF calculations overestimate $k_F$ or $1/R_c$, which implies that the predicted average density of the nuclear systems is too large. With respect to the energy per nucleon, however, the BHF calculations provide too much energy in nuclear matter and too little for the finite nucleus. Therefore, to improve the comparison with experiment, the inclusion of the same 3NF must provide attraction in finite nuclei and repulsion in infinite matter and reduce the calculated saturation density in both cases.

The inclusion of rearrangement terms going from the  BHF to the RBHF approach yields occupation probabilities $P_i$ of the order of 0.8 to 0.9 as shown in  Table~\ref{VergleichO1}. From Eq.~(\ref{eq:rbhf1}) it is obvious that this leads to less attractive single-particle energies in RBHF as compared to the BHF approach. The weaker attraction of the single-particle potential is reflected in a larger radius of the charge distribution. On the other hand the less attractive single-particle energies yield less attractive starting energies $\omega$ in the Bethe-Goldstone equation, which leads to more attractive $G$-matrix elements and thus larger binding energy per nucleon. This means that the inclusion of rearrangement terms shifts both the energy per nucleon and the charge radius closer to their experimental values. As one can see from Table~\ref{VergleichO1} and  Fig.~\ref{fig:oca1}, this effect is too small to provide a satisfactory agreement.

The nuclear matter study discussed above already showed that the results of BHF calculations are rather sensitive to a consistent treatment of the two-particle propagator in the Bethe-Goldstone equation. The treatment of the particle-state spectrum, in particular, requires special attention. The same is of course to be expected also for finite nuclei. Therefore Lippok and one of us\cite{lippok} considered a reasonable variation of the specrum for the particle states. In this figure, the results of BHF calculations with different particle-state spectra are represented by symbols connected by a solid red line, while the symbols connected by a dashed blue line refer to the results of RBHF calculations.
The remarkable sensitivity of the calculation to these changes in the particle-state spectrum calls for more sophisticated investigations of this issue in order to obtain unambiguous results.

The effects of including the 3NF can be seen by comparing the results shown in Fig.~\ref{fig:oca1} by square boxes with the corresponding results displayed by circles. As to be expected, the inclusion of the 3NF yields a repulsive effect and leads to a reduction of the total energy accompanied by an increase of the nuclear radius. Comparing this effect with the corresponding repulsive effect one can obtain with the change in the particle-state spectrum just discussed, one finds that the 3NF yields a larger increase in the radius, if the energy is changed by a similar amount. 

As can be seen from the right panel of  Fig.~\ref{fig:oca1}, the main features of the results for $^{40}$Ca are very similar to those found in $^{16}$O. Therefore, one may conclude that using an appropriate 3NF to simulate the effects of Dirac spinors modified in the nuclear medium within the framework of non-relativistic BHF calculations provides a good description of the bulk properties of nuclear matter and finite nuclei based on a realistic $NN$ interaction.

The description of bulk properties (energy, radius of particle distribution, density) is an important but only one of features of nuclear structure which one hopes to describe within the relativistic DBHF. Other important aspects are the energy dependence of the optical potential (see Section~\ref{sect4}) and the spin-orbit splitting of the single-particle energies,\cite{Zamick} which is enhanced due to the enhancement of the small component of Dirac spinors in the nuclear medium (see discussion of Eq.~(\ref{eq15})).

Can this enhancement of the spin-orbit splitting (which is crucial to describe the strength of the spin-orbit term observed in the experiment), be simulated by the simple 3NF of Eq.~(\ref{eq:3-vp})? Inspecting e.g. the single-particle energies of the $p_{3/2}$ and $p_{1/2}$ states listed in Table~\ref{VergleichO1}, we do not find an enhancement of the spin-orbit splitting when the 3NF is included. In fact, the differences between these single-particle energies are always smaller with inclusion of the 3NF. This is related to the fact that the 3NF yields larger values for the radii which, in turn, reduces the spacing between the single-particle states. But even if one accounts for this size effect, the 3NF force does not provide an enhancement of the spin-orbit splitting. This could be achieved by introducing an appropriate spin structure in the 3NF, a feature which presumably would lead to more parameters and diminish the simplicity of the present approach.

\section{Summary and conclusions \label{sect6}} 

In this article, we presented a broad survey of relativistic effects in nuclear matter and 
nuclei from the point of view of a microscopic approach. 

First, we reviewed the general aspects of the Dirac-Brueckner-Hartree-Fock approach as 
applied to infinite nuclear matter. We recall how 
the DBHF main feature, namely a characteristic enhancement of the lower component of the 
nucleon spinor in the medium, originates from the most general Lorentz structure of the single-particle
operator, or self-energy. This mechanism, known as the ``Dirac effect", provides the additional
repulsion, as compared to the conventional (non-relativistic) approach, necessary to predict a realistic       
behaviour of the nuclear matter equation of state at saturation. 

We extended our discussion to infinite matter with different concentrations of protons and 
neutrons. This brings up the symmetry energy, a topic of 
contemporary interest, particularly as new experimental facilities and programs promise 
to improve our knowledge of neutron-rich systems.                                                    
At this point in the article, we noted that                    
chiral EFT has become very
popular in recent years as a way to respect the symmetries of 
low-energy QCD while retaining the degrees of freedom typical 
of nuclear physics, nucleons and pions. We then proceeded to a comparison 
between two very different methods, both microscopic,                        
to approach the study of nucleonic matter: one based on  meson-theoretic relativistic 
potentials and the DBHF approximation; the other based on a high-precision
chiral $NN$ potential and the leading chiral 3NF.                           

Both approaches lead to rather similar results for infinite matter. They reproduce the empirical saturation point of nuclear matter combining the quenching
of two-nucleon correlation included in the Brueckner-Hartree-Fock approach with a repulsive 3NF, if we interpret the ``Dirac effect'' also as a kind of
3NF. As the chiral EFT leads to a softer $NN$ interaction than typical meson-theoretical models, a much stronger 3NF is required in the former than in
the letter. This suggests that the study of correlation effects might be usefull to distinguish between these two approaches.
                                        
After surveying various results,                                                                                         
we argue that the DBHF approach is a valid paradigm, particularly 
when high momenta (and/or high Fermi momenta) are involved, as those are unaccessible to chiral EFT.                     

A central part of this review has been devoted to a discussion of the DBHF approach as applied to 
finite nuclei. We reviewed the Optical Model Potential based on DBHF for nucleon-nucleus 
scattering, which is crucial for the derivation of many reaction observables.
We conclude that the DBHF approach is able to provide detailed information on nucleon-nucleus observables
without the need to adjust a large number of parameters, a procedure which is
necessary in the determination of phenomenological 
global optical potentials.

A focal point of our discussion has been the interpretation of the Dirac effects, namely
relativistic density-dependent effects, in terms of an effective 3NF. One finds that a good description
of bulk properties of both nuclear matter and finite nuclei can be achieved with a realistic
$NN$ interaction and an appropriate 3NF            
to simulate Dirac effects. There are, of course, limitations to the 
predictive power of this simple prescription, especially when spin structures are involved, in particular the
strength of the spin-orbit splitting in the single-particle energies. 
Overall, the ability of the DBHF approach to incorporate important three-body effects 
continues to make this method attractive and convenient.                  

\section*{Acknowledgements}
Support from the U.S. Department of Energy under grant DE-FG02-03ER41270 and the Deutsche Forschungsgemeinschaft (DFG, contract Mu 705/10-1)
is acknowledged.

\end{document}